\documentclass{jfm}

\usepackage{graphicx, subfig}
\usepackage{natbib, srctex}
\usepackage{amsmath, amssymb, wasysym}

\ifCUPmtlplainloaded \else
  \checkfont{eurm10}
  \iffontfound
    \IfFileExists{upmath.sty}
      {\typeout{^^JFound AMS Euler Roman fonts on the system,
                   using the 'upmath' package.^^J}%
       \usepackage{upmath}}
      {\typeout{^^JFound AMS Euler Roman fonts on the system, but you
                   dont seem to have the}%
       \typeout{'upmath' package installed. JFM.cls can take advantage
                 of these fonts,^^Jif you use 'upmath' package.^^J}%
      }
  \else
  \fi
\fi

\ifCUPmtlplainloaded \else
  \checkfont{msam10}
  \iffontfound
    \IfFileExists{amssymb.sty}
      {\typeout{^^JFound AMS Symbol fonts on the system, using the
                'amssymb' package.^^J}%
       \usepackage{amssymb}%
       \let\le=\leqslant     
      }{}
  \fi
\fi

\ifCUPmtlplainloaded \else
  \IfFileExists{amsbsy.sty}
    {\typeout{^^JFound the 'amsbsy' package on the system, using it.^^J}%
     \usepackage{amsbsy}}
    {\providecommand\boldsymbol[1]{\mbox{\boldmath $##1$}}}
\fi

\newsavebox{\astrutbox}
\sbox{\astrutbox}{\rule[-5pt]{0pt}{20pt}}

\newcommand{\tensor}[1]{\overline{\overline{\mathbf{#1}}}}

\title[Drag-adjoint of cylinder wake at $Re_D=20$, 100 and 500]
{The drag-adjoint field of a circular cylinder wake at Reynolds
numbers 20, 100 and 500}

\author[Q. Wang and J.-H. Gao]
{QIQI WANG$^1$ \and JUN-HUI GAO$^2$}

\affiliation{
$^1$Department of Aeronautics and Astronautics, MIT, Cambridge, MA, US
\\[\affilskip]
$^2$Beihang University, Beijing, China
}

\pubyear{2012}
\volume{538}
\pagerange{1--26}
\date{?? and in revised form ??}
\begin{document}

\maketitle

\begin{abstract}
This paper analyzes the adjoint solution of the Navier-Stokes equation.
We focus on flow across a circular cylinder at three Reynolds numbers,
$Re_D=20, 100$ and $500$.  The quantity of interest in the adjoint
formulation is the drag on the cylinder.
We use classical fluid mechanics approaches to analyze the adjoint
solution, which is a vector field similar to a flow field.
Production and dissipation
of kinetic energy of the adjoint field is discussed.  We also
derive the evolution of circulation of the adjoint field along a closed
material contour.  These analytical
results are used to explain three numerical solutions of the adjoint
equations presented in this paper.  The adjoint solution at $Re_D=20$,
a viscous steady state
flow, exhibits a downstream suction and an upstream jet, the opposite
of the expected behavior of a flow field.  The adjoint solution at
$Re_D=100$, a periodic 2D unsteady flow, exhibits periodic,
bean-shaped circulation in the near-wake region.  The adjoint solution at
$Re_D=500$, a turbulent 3D unsteady flow, has complex dynamics created
by the shear layer in the near wake.  The magnitude of the adjoint
solution increases exponentially at the rate of the first Lyapunov
exponent.  These numerical results correlate well with the theoretical
analysis presented in this paper.
\end{abstract}


\section{Introduction}
\label{s:intro}

The adjoint method for sensitivity analysis has been a powerful tool
in computational fluid dynamics for over 20 years
\cite[]{Jameson1988, Becker_2001_An_Optimal_Control_Approach,
Pierce_2000_Adjoint_Recovery_of_Superconvergent_Functionals}.
This method enables
efficient computation of sensitivity gradients in fluid flow problems,
and is widely used in aerodynamic shape optimization, inverse design
problems, adaptive mesh refinement and uncertainty quantification.

Traditionally, the adjoint field is
regarded mainly as a mathematical quantity, representing the sensitivity
derivative of a quantity of interest to residual of the Navier-Stokes
equation.
Recently, \cite{cylinderAdjointSens} explored physical aspect of the
adjoint velocity field to study the stability of a cylinder wake
at $Re_D=46.8$.
This paper further explores physical aspects of the adjoint
field across three different Reynolds number regimes.
We focus on the quantity of interest in the adjoint formulation,
which is the drag on an object in flow.  This
quantity of interest defines a particular boundary condition for the
adjoint equation, under which the adjoint solution is called the drag-adjoint
field.  We show that the drag-adjoint field is a
non-dimensional physical quantity; it represents the transfer function
from small forces applied to the fluid flow to drag on the object in the
fluid flow.

The drag-adjoint field is analyzed for flow across a circular
cylinder at three different Reynolds numbers, 20, 100 and 500, based on
the freestream velocity and the cylinder diameter.
These Reynolds numbers fall into the laminar steady regime, laminar
vortex shedding regime, and the disordered three-dimensional regime,
respectively.  These regimes are characterized by \cite{cylinder} for
the circular cylinder wake, and are present in many bluff body wakes.
At $Re_D=20$, the flow field is steady and stable.
At $Re_D\approx 46$, the vortex
wake undergoes a supercritical Hopf bifurcation into a periodic limit
cycle oscillation
\cite[]{cylinderHopf,cylinderSens,cylinderBenardVonKarmanHopf},
which further develops into von Karman vortex shedding.
The drag-adjoint field of such periodic vortex shedding is
studied at $Re_D=100$ in this paper.  As the Reynolds number further
increases above $Re_D\approx 189$, streamwise vortices develop,
and the wake goes through a series
of transitions to turbulence.  Studies of this transition are
summarized by \cite{cylinder}.
In this work, the drag-adjoint field of turbulent wake structure at $Re_D=500$
is studied.

Flow across a circular cylinder at $Re_D=20$, 100 and 500 represents three
distinct types of dynamical system:
At $Re_D=20$, the steady state flow is
a fixed point attractor in the state space.  If the flow field is
perturbed from the steady state, it will relax to the steady state as
time advances -- in other words, all
Lyapunov exponents of this dynamical system are negative.  At $Re_D=100$, the
periodic van Karman vortex shedding represents a limit cycle attractor in the
state space.  If the flow field is perturbed, it will relax to the same
periodic oscillation, with a potential phase difference.  In other
words, the maximal Lyapunov exponent is zero.  At $Re_D=500$,
the fluid flow is chaotic.  The aperiodic oscillations in the turbulent wake
represent a strange attractor in the state space.
The system is chaotic and has at least one positive Lyapunov exponent.

The rest of this paper is organized as follows: section \ref{s:analysis}
starts by using classic fluid mechanical methods to perform theoretical
analysis of the behavior of the adjoint field;
section \ref{s:computation} analyzes numerical solutions of the
adjoint equation at three Reynolds numbers, and correlates the results
with theoretical analysis; and section \ref{s:conclusion} concludes this
paper.

\section{Theoretical Analysis of the Drag-Adjoint Field}
\label{s:analysis}

\subsection{Mathematical basis of the drag-adjoint}

We consider a circular cylinder in a freestream
$\mathbf{u}_{\infty} = (1,0,0)$.  The fluid flow field is governed by the
incompressible Navier-Stokes equation with constant
density $\rho$ and viscosity $\mu$:
\begin{equation} \label{nseqn}
 \rho \frac{\partial \mathbf{u}}{\partial t} + \rho\, \mathbf{u}\cdot \nabla \mathbf{u} + \nabla p
 = \mu \nabla \cdot \nabla \mathbf{u} \;,\quad
 \nabla\cdot \mathbf{u} = 0 
\end{equation}
with boundary condition $\mathbf{u}=(0,0,0)$ at the cylinder surface $S$.
The drag-adjoint field $\hat{\mathbf{u}}$ is a nondimensional vector
field that satisfies the adjoint equation
\begin{equation} \label{adjoint}
 \rho \frac{\partial \hat{\mathbf{u}}}{\partial t} + \rho\,\mathbf{u}\cdot \nabla \hat{\mathbf{u}}
 - \rho\,\nabla \mathbf{u} \cdot \hat{\mathbf{u}} + \nabla \,\hat{p}
 = -\mu \nabla \cdot \nabla \hat{\mathbf{u}} \;,\quad
 \nabla\cdot \hat{\mathbf{u}} = 0
\end{equation}
with boundary condition $\hat{\mathbf{u}}=(1,0,0)$ at the cylinder surface $S$,
and $\hat{\mathbf{u}}=(0,0,0)$ in the far field.
Note that the adjoint equation should be viewed as evolving backwards in time;
the negative viscosity in the right hand side is a dissipative term
in this sense.  Equation (\ref{adjoint}) and its boundary conditions
are crafted to be adjoint to the linearized Navier-Stokes equation
via integration by parts.  In particular, the boundary conditions arise
from the bilinear concomitant after the integration by parts.
More details are given in the Appendix \ref{s:app}.

Mathematically, the drag-adjoint field represents the $L^2$
functional derivative of the time-averaged drag force to body forces in the
fluid flow field.  It can be derived by analyzing the effect of a small
perturbation to the drag
\begin{equation}
D = \oiint_S \left(p\,n_x - \mu \mathbf{n} \cdot \nabla u_x\right) ds\;.
\end{equation}
where the surface normal $n$ points from the fluid domain into the
cylinder.
The perturbation in $D$ due to an infinitesimal body force
$\delta\mathbf{f}$ in the
interior of the fluid flow field is
\begin{equation} \label{deltaD}
\delta D = \oiint_S \left(\delta p\,n_x
- \mu \mathbf{n} \cdot \nabla \delta u_x\right) ds\;.
\end{equation}
where $\delta p$ and $\delta \mathbf{u}$ are infinitesimal perturbations of the
fluid flow, governed by the tangent linear Navier-Stokes equation,
with the infinitesimal body force as an additional right hand side:
\begin{equation} \label{linearized}
 \rho \frac{\partial \delta \mathbf{u}}{\partial t} + \rho\, \mathbf{u}\cdot \nabla \delta \mathbf{u}
 + \rho\,\delta  \mathbf{u}\cdot \nabla \mathbf{u} + \nabla \delta p
 = \mu \nabla \cdot \nabla \delta \mathbf{u} + \delta\mathbf{f} \;,\quad
 \nabla\cdot \delta \mathbf{u} = 0 
\end{equation}
Multiply $\hat{\mathbf{u}}$ onto the tangent linear momentum equation, and
integrate over the fluid domain (details in the Appendix \ref{s:app}).
By applying integration by parts
and the drag-adjoint equation, we obtain
\begin{equation} \label{unsteady}
 \rho\, \frac{d}{dt} \iiint \hat{\mathbf{u}}\cdot\delta \mathbf{u}\: dV
 + \oiint_S \hat{\mathbf{u}}\cdot (\delta p\,\mathbf{n} - \mu \mathbf{n} \cdot \nabla \delta u_x)\:ds
 = \iiint \hat{\mathbf{u}}\cdot \delta\mathbf{f}\:dV
\end{equation}
Because of the adjoint boundary condition, $\hat{\mathbf{u}}=(1,0,0)$ on
the cylinder surface $S$; therefore, the second term in (\ref{unsteady})
is $\delta D$ in Equation (\ref{deltaD}).
By integrating the equality over time, we thus obtain
\begin{equation} \label{unsteady2}
  \left.\rho\, \frac{d}{dt} \iiint \hat{\mathbf{u}}\cdot\delta \mathbf{u}\: dV \right|_0^T
+ \int_0^T \delta D\: dt
= \int_0^T \iiint \hat{\mathbf{u}}\cdot \delta\mathbf{f}\:dV dt
\end{equation}
In particular, if $\delta \mathbf{u}=0$ at $t=0$ and $\hat{\mathbf{u}}=0$ at $t=T$, we
have
\begin{equation} \label{steady}
  \int_0^T \delta D\: dt
= \int_0^T \iiint \hat{\mathbf{u}}\cdot \delta\mathbf{f}\:dV dt
\end{equation}
In other words, the small perturbation in the time-averaged drag is equal to
the time-averaged integral of the inner product between the adjoint vector
and the perturbing force.

\subsection{The adjoint field as a nondimensional transfer function}

Equation (\ref{steady}) reveals the physical meaning of the
drag-adjoint field $\hat{\mathbf{u}}$.  It is a non-dimensional transfer
function from forces applied in the fluid to drag force on the cylinder.
To clearly illustrate this concept, imagine $\delta\mathbf{f}$ to be a Dirac delta
function at time $t$ and spatial point $x$ inside the fluid domain of magnitude
$\boldsymbol{\epsilon}$.
This represents an infinitesimal impulse applied to the fluid flow
at time $t$, concentrated at the point $x$.  The resulting impact on the
cylinder, according to (\ref{steady}), is
\begin{equation}
 \int_0^T \delta D\,dt = \hat{\mathbf{u}}(x, t)\cdot \boldsymbol\epsilon
\end{equation}

The value $\hat{\mathbf{u}}$ at $x$ reveals how a small impulse at $x$ and $t$
is transferred to the cylinder as drag.  In particular, if the impulse
$\boldsymbol{\epsilon}$ is along the direction of the drag-adjoint field $\hat{\mathbf{u}}$,
it increases the mean drag on the cylinder; if the impulse is against the
direction
of $\hat{\mathbf{u}}$, it decreases the drag on the cylinder; if the impulse is
perpendicular to $\hat{\mathbf{u}}$, it has no (first order) effect on the
drag of the cylinder.

The magnitude of the drag-adjoint field $\hat{\mathbf{u}}$ represents
what fraction of a small impulse
$\boldsymbol{\epsilon}$ in the fluid domain is transferred as drag on the cylinder
surface.  If the magnitude of the drag-adjoint field $\hat{\mathbf{u}}$ is
smaller than 1, then a small impulse of $\boldsymbol{\epsilon}$ Newton second along the
direction of $\hat{\mathbf{u}}$ results in a less than $\boldsymbol{\epsilon}$
Newton second increase
in the time-integrated cylinder drag; a larger than 1 magnitude means that a
small impulse
along the direction of the drag-adjoint field is amplified through
the dynamics of the flow, and applied as drag on the cylinder.

This physical interpretation of the drag-adjoint field enables us
to examine the drag-adjoint field as a physical field, as well
as a functional derivative in a mathematical sense.  We believe
that by observing the behavior of the drag-adjoint field, one could
gain additional physical insight into the dynamics of the fluid flow
that would be difficult to learn from observing the fluid flow itself.

\subsection{Energy balance of the adjoint field}

The $L^2$ energy of the adjoint field
\begin{equation}
\hat{\mathcal{E}} = \iiint \frac12 \rho\, \hat{\mathbf{u}}\cdot\hat{\mathbf{u}}\,dV
\end{equation}
is an integral measure of sensitivity.
It provides a tight upper bound of how sensitive the drag
is with respect to small perturbations applied to the flow field.
Applying the Cauchy-Schwarz inequality to Equation (\ref{steady}),
one obtains
\begin{equation}
  \left| \int_0^T \delta D\: dt \right|
\le \int_0^T \left(\frac{2 \hat{\mathcal{E}}}{\rho}
            \iiint \delta\mathbf{f} \cdot \delta\mathbf{f}\,dV\right)^{\frac12} dt\;,
\end{equation}
where equality holds when $\delta\mathbf{f}\propto \hat{\mathbf{u}}$.

The evolution of $\hat{\mathcal{E}}$
can be analyzed by multiplying the adjoint equation (\ref{adjoint}) with
$\hat{\mathbf{u}}$.  Through integration by parts, we obtain
\begin{equation} \label{energy} \begin{split}
-\frac{d}{dt}\hat{\mathcal{E}} &=
 \iiint -\rho\hat{\mathbf{u}}\cdot\nabla \mathbf{u}\cdot\hat{\mathbf{u}}\,dV
+\iiint -\mu \|\nabla\hat{\mathbf{u}}\|^2\, dV + \hat{P}^{BC} \\
                               &=
 \iiint \hat{P}^{\hat{\mathcal{E}}} \, dV -
 \iiint \hat{D}^{\hat{\mathcal{E}}} \, dV + \hat{P}^{BC}
\end{split} \end{equation}
where $\hat{P}^{BC}$
represents the boundary terms resulting from integration by
parts; $\|\cdot\|$ is the Frobenius norm of the vector field gradient tensor.
The negative sign on the left side is because the adjoint field evolves
backwards in time.

To simplify the boundary
term $\hat{P}^{BC}$ resulting from integration by parts, we use
both the fluid flow boundary condition and the adjoint boundary
condition, i.e.,
$\mathbf{u}=(0,0,0),\hat{\mathbf{u}}=(1,0,0)$ on the cylinder surface $S$ and
$\mathbf{u}=(1,0,0),\hat{\mathbf{u}}=(0,0,0)$ on the far field $F$:
\begin{equation} \label{eneprodbc} \begin{split}
\hat{P}^{BC} &= \oiint_{S\cup F}
         \left(\frac{\hat{\mathbf{u}} \cdot\hat{\mathbf{u}}}{2}\,\mathbf{u} + \hat{p}\,\hat{\mathbf{u}}
             + \mu \nabla\hat{\mathbf{u}} \cdot \hat{\mathbf{u}}\right)\cdot \mathbf{n}\, ds \\
       &= \oiint_{S} \left(\hat{p} + \nabla\hat{\mathbf{u}}\right) n_x \, ds
\end{split}\end{equation}
Note that the boundary term $\hat{P}^{BC}$ is proportional to
the magnitude of the
adjoint solution $\hat{\mathbf{u}}$ and $\hat{p}$, while the other terms
$\hat{P}^{\hat{\mathcal{E}}}$ and $\hat{D}^{\hat{\mathcal{E}}}$
in Equation (\ref{energy}) are quadratic with respect to the magnitude
of the adjoint field.

The second term in Equation (\ref{energy}),
$\hat{D}^{\hat{\mathcal{E}}}=\mu \|\nabla\hat{\mathbf{u}}\|^2$,
is clearly a dissipation term.
It is always negative, thus always takes energy away from the adjoint
field as the adjoint evolves backwards
in time.  The rate of energy dissipation is proportional to viscosity
and the gradient of the adjoint.
The first term in Equation (\ref{energy}) is the ``production''
of adjoint energy in the interior of the fluid domain.
Here we take a more detailed look at the pointwise contribution
to the production of adjoint energy
$\hat{P}^{\hat{\mathcal{E}}} = -\rho\, \hat{\mathbf{u}}\cdot\nabla \mathbf{u}\cdot\hat{\mathbf{u}}$.

We first observe that the
anti-symmetric part of the velocity gradient tensor $\nabla \mathbf{u}$ does
not contribute to the energy generation.  This is because an
anti-symmetric tensor is diagonalizable by an orthonormal transform and
contains only zero and pure imaginary eigenvalues.  In other words,
vorticity \emph{per se} does not contribute to the production of the adjoint.
The only contribution to the production of the adjoint field is
the symmetric part of the velocity gradient tensor $\nabla \mathbf{u}$, i.e.
the rate of shear strain $\tensor{S} = \frac12(\nabla \mathbf{u} + \nabla \mathbf{u}^T)$.
The production of adjoint energy can therefore be rewritten as
\begin{equation}\label{adjeneprod}
\hat{P}^{\hat{\mathcal{E}}} = -\rho\, \hat{\mathbf{u}}\cdot \tensor{S}\cdot\hat{\mathbf{u}}.
\end{equation}

In addition, one can decompose the shear strain rate into its principal
components
\begin{equation}
\tensor{S} = \lambda_1 \:\mathbf{q}_1 \otimes \mathbf{q}_1
           + \lambda_2 \:\mathbf{q}_2 \otimes \mathbf{q}_2
           + \lambda_3 \:\mathbf{q}_3 \otimes \mathbf{q}_3
\end{equation}
For incompressible flow, the tensor has zero trace, therefore $\lambda_1
+ \lambda_2 + \lambda_3=0$.  The principal components corresponding to
negative eigenvalues are the contracting directions, and contribute to
positive energy production in Equation (\ref{adjeneprod}); the principal
components corresponding to positive eigenvalues are the stretching
directions, and contribute to negative energy production.
The sign of the combined production from all principal components
depends on the direction of the adjoint vector $\hat{\mathbf{u}}$. If it is
aligned with the principal components with negative eigenvalues (the
contracting directions forwards in time and stretching directions
backwards in time), the production of adjoint energy is positive.
If $\hat{\mathbf{u}}$ is aligned with the principal components with positive
eigenvalues (the stretching directions forwards in time and contracting
directions backwards in time), the production of adjoint energy is
negative.

In the adjoint energy equation (\ref{energy}),
the relative magnitude of the net contribution from the production term
$\hat{P}^{\hat{\mathcal{E}}} = -\rho\, \hat{\mathbf{u}}\cdot \tensor{S}\cdot\hat{\mathbf{u}}$
to the net dissipation from
$\hat{D}^{\hat{\mathcal{E}}}=\mu \|\nabla\hat{\mathbf{u}}\|^2$ determines
the energy balance of the adjoint field.
When $\hat{P}^{\hat{\mathcal{E}}}$ is smaller than
$\hat{D}^{\hat{\mathcal{E}}}$, the main effect in the interior of the
domain is damping; most of the energy of the adjoint field
comes from the boundary terms in the Equation (\ref{energy}).
This is common for steady, laminar flows.
When $\hat{P}^{\hat{\mathcal{E}}}$
and $\hat{D}^{\hat{\mathcal{E}}}$ have similar magnitudes, the adjoint
field could produce enough energy in the interior of domain to sustain itself.
This is common for periodic, laminar flows.
When $\hat{P}^{\hat{\mathcal{E}}}$ is greater than
$\hat{D}^{\hat{\mathcal{E}}}$, the adjoint field can self-produce
net positive energy from the interior of the domain, leading to
exponential growth of the adjoint solution.  This is common for
chaotic, turbulent flows.  Adjoint fields for flow over a circular cylinder at
$Re_D=20, 100$ and $500$ correspond to these three cases, respectively.

\subsection{Circulation dynamics of the adjoint field}
\label{s:circulation}

Being a solution of the adjoint equation (\ref{adjoint}), the
drag-adjoint field $\hat{\mathbf{u}}$ is divergence-free.  Thus it can be
characterized by its vorticity field $\nabla\times\hat{\mathbf{u}}$.
As will be shown in Section \ref{s:computation}, the 
drag-adjoint field of a circular cylinder wake is often dominated by
eddy-like structures.  These observations motivate us to use classic
fluid mechanics techniques to analyze the behavior of circulation of
the adjoint field, which by the Stokes theorem is an integral measure
of vorticity.  Note that this
analysis does not depend on the specific boundary condition of the adjoint
field, and is therefore applicable to adjoint fields of many other quantities
of interest.  To begin this analysis, we study the evolution of circulation
of the adjoint field around a fluid contour $C$
\begin{equation}
\hat\Gamma_C = \oint_C \hat{\mathbf{u}} \cdot d\mathbf{l}
\end{equation}
Because the adjoint equation is viewed as evolving backwards in time,
we analyze the negative of the Lagrangian derivative of the adjoint
circulation,
\begin{equation}
-\frac{D}{Dt}\hat\Gamma_C = -\oint_C \frac{D\hat{\mathbf{u}}}{Dt} \cdot
d\mathbf{l}
                           - \oint_C \hat{\mathbf{u}} \cdot
                             \frac{D\,d\mathbf{l}}{Dt}
\end{equation}
In this equation,
the Lagrangian time derivative of $\hat{\mathbf{u}}$ is determined by the
adjoint equation (\ref{adjoint}).
Also, the rate of stretching of an infinitesimal fluid line $d\mathbf{l}$
is equal to the velocity difference between the two end points of the
fluid line $d\mathbf{l}$, i.e., $\frac{D d\mathbf{l}}{Dt} = d\mathbf{u}$.  With these
substitutions, we obtain
\begin{equation}
-\frac{D}{Dt}\hat\Gamma_C = -2 \oint_C \hat{\mathbf{u}} \cdot d\mathbf{u}
                         + \mu \oint_C \nabla \cdot \nabla
                         \hat{\mathbf{u}}\cdot d\mathbf{l}
\end{equation}

Evolving backwards in time, the second term in this equation is
equivalent to the effect of
viscosity on fluid velocity circulation.  However, the first term is
unique for the adjoint circulation.
Here we analyze the physical meaning of this term, namely the
production of adjoint circulation
\begin{equation} \label{productadj}
P^{\Gamma}_C  = -2 \oint_C \hat{\mathbf{u}} \cdot d\mathbf{u}
= -2 \oint_C \hat{\mathbf{u}} \cdot \frac{D\,d\mathbf{l}}{Dt}
\end{equation}
With Stokes theorem, it can also be written as
\begin{equation} \label{productadjstokes}
P^{\Gamma}_C
= -2 \oint_C \hat{\mathbf{u}} \cdot \nabla \mathbf{u} \cdot d\mathbf{l}
= -2 \iint_{\Sigma_C} \left(\nabla \hat{u}_1 \times \nabla u_1 
                 + \nabla \hat{u}_2 \times \nabla u_2
                 + \nabla \hat{u}_3 \times \nabla u_3\right) \cdot
                 d\mathbf{s}
\end{equation}
where $\Sigma_C$ is a surface enclosed by the curve $C$, and
$d\mathbf{l}$ points anticlockwise when the surface normal
$d\mathbf{s}$ points toward the viewer, following the right-hand rule.
We make two main observations on the production of adjoint circulation:
\begin{itemize}
\item
The production of adjoint circulation is an
inviscous effect.  It has no dependence on viscosity.  Therefore,
Kelvin's theorem does not hold for the adjoint
field, even in the absence of viscosity.  
\item According to Equation (\ref{productadj}),
amplification of adjoint circulation happens when the
adjoint vector $\hat{\mathbf{u}}$ is aligned against the rate of stretching of the
fluid flow $\frac{D\,d\mathbf{l}}{Dt}$.  In other words,
when two fluid particles along the direction of the
adjoint vector are squeezed against each other.   Adjoint circulation
is removed when the rate of stretching of the fluid flow is in the same
direction as the adjoint vector field, i.e. when two fluid particles
along the direction of the adjoint vector moves away from each other.
Another point of view is that an adjoint eddy is amplified when
it is "stretched" backwards in time by the fluid flow;
an adjoint eddy is damped when it is "squeezed"
backwards in time by the fluid flow.
\end{itemize}

\begin{figure}\centering
  \subfloat[Elongated adjoint ring in shear flow]{\label{elongated}
      \includegraphics[width=0.3\textwidth]{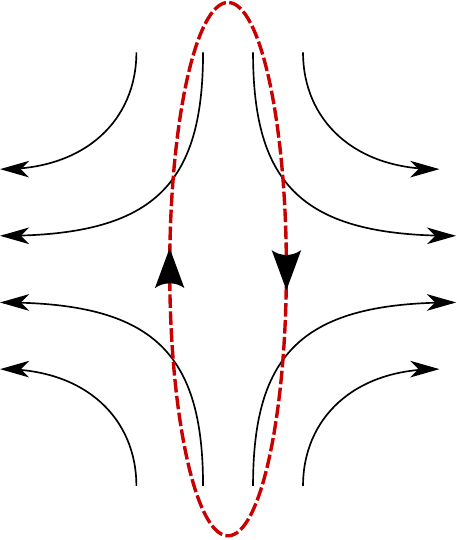}}
  \hspace{0.03\textwidth}
  \subfloat[Elongated rings in parallel shear layer]{\label{shearadj}
      \includegraphics[width=0.3\textwidth]{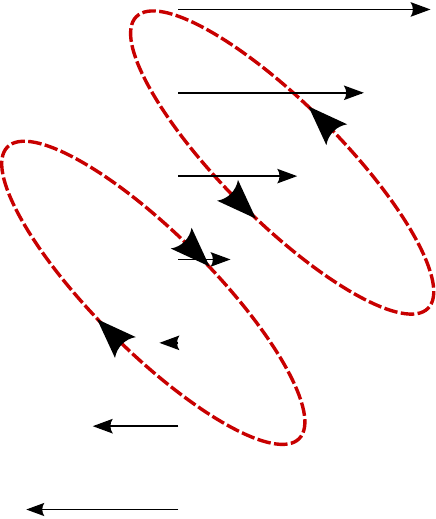}}
  \hspace{0.03\textwidth}
  \subfloat[Bean shaped ring in a wake]{\label{wakeadj}
      \includegraphics[width=0.3\textwidth]{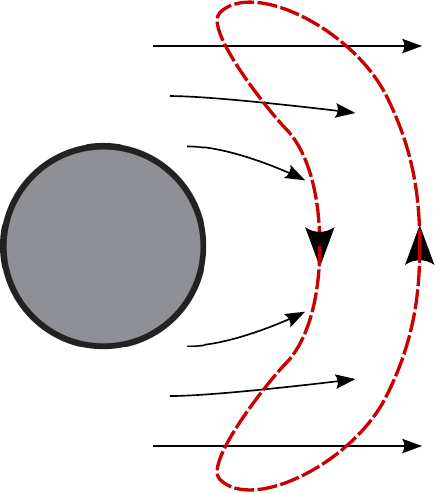}}
  \caption{Amplification of elongated adjoint circulation}
\label{adjointcirc}
\end{figure}

The second observation can explain some features of the adjoint fields
we observe in the cylinder wake.  In particular, a commonly observed
structure of the adjoint field is elongated eddies,
which often happens in shear flows.  Figure \ref{elongated} illustrates
how such eddies are preferentially amplified in shear flow.
In this figure, the thin black solid lines represent the fluid
velocity, and the thick red dashed lines represent an elongated
eddy of the adjoint field.  In this flow, fluid particles along
 the vertical direction are squeezed
against each other, while fluid particles along the horizontal direction
are stretched.
Now consider the production of circulation along the elongated adjoint
eddies, as defined in Equation (\ref{productadj}).  The long, vertical
legs of the eddy experience squeezing forwards in time (stretching backwards
in time).  Therefore, they contribute to amplification of circulation
according to (\ref{productadj}).  The
short, horizontal legs of the elongated adjoint eddy experience stretching
forwards in time (squeezing backwards in time).  They contribute to damping
of the adjoint circulation according to (\ref{productadj}).
Because the horizontal legs are shorter than the vertical
legs, the main effect is
amplification of adjoint circulation of such adjoint eddies,
elongated along the convergent direction of the shear flow.

As an example of the analysis above,
consider a parallel shear layer as illustrated in Figure
\ref{shearadj}.  The flow velocity gradient in the parallel shear layer
consists of clockwise rotation (anti-symmetric part of the velocity
strain tensor)
and a symmetric shear in the 45 degrees direction (symmetric part of the
velocity strain tensor).  As a result, the adjoint circulation aligned
45 degrees to the
shear layer is the most amplified, as shown in Figure \ref{shearadj}.
It is worth noting that this amplification effect is non-directional,
i.e., it equally amplifies elongated adjoint rings with either positive
or negative circulation.  Therefore, multiple counter-rotating,
parallel elongated adjoint rings can be produced in the same shear layer,
as illustrated in Figure \ref{shearadj}.

The elongated adjoint circulation in shear flows, as illustrated in
Figure \ref{adjointcirc}, is observed in the adjoint field of cylinder
wakes.  These adjoint flow patterns can be related to Kelvin-Helmholtz modes
in shear flows, because they represent external force patterns to the
fluid that have most influence on the flow field, and ultimately the
drag on the cylinder.  While Kelvin-Helmholtz is a 2D phenomenon,
more types of instabilities could occur in three dimensions, potentially causing
more patterns to appear in the adjoint field.

It is worth noting that amplification of a pattern in the adjoint field
may indicate either modal or non-modal (i.e., transient) growth of
perturbations in the flow field (\cite{nonmodal}).
In fact, the adjoint field has been
a very effective tool for studying non-modal stability of unsteady
flows.  The combination of modal and non-modal modes makes the adjoint
field very complex, especially in 3D flows.

\section{Computational results}
\label{s:computation}

The results presented in this section are obtained by discretizing
both the Navier-Stokes equation (\ref{nseqn}) and the continuous form
of the adjoint equation (\ref{adjoint}) with a second order finite volume
scheme \cite[]{cdp1, qiqiwang_thesis}.
The cylinder diameter is 1; the freestream velocity is
$(1,0,0)$; the Reynolds number is set by the viscosity $\nu=1/Re_D$.
A 2D mesh of about $23,500$ control volumes is used for the $Re_D=20$
and $Re_D=100$ cases.  The $Re_D=500$ case is solved with a spanwise
extent of 4 cylinder diameters. A total number of 2.06 million
control volumes are used for the 3D simulation.

For each case,
the Navier-Stokes equation is first solved for sufficiently
long time to reach a steady or quasi-steady state at $T_0$.  The Navier-Stokes
equation is then further integrated forwards in time to $T_1$; the fluid
flow field $\mathbf{u}$ between $T_0$ and $T_1$ is stored in a dynamic checkpointing
scheme \cite[]{wang_siam08}.  The adjoint field is initialized at $T_1$
to $\hat{\mathbf{u}}(x, T_1) = (0,0,0)$, and integrated backwards in time to
$T_0$.  During the backwards time integration, the fluid flow field $\mathbf{u}$
required by the adjoint integrator is retrieved from the checkpointing
scheme.  The time length of the adjoint solution $T_1-T_0$ is 50 for
all three cases.

\subsection{Adjoint flow field of steady wake at $Re_D=20$}

At Reynolds number $Re_D=20$, the fluid flow is steady and stable.
Any small perturbation to the fluid flow field asymptotically decays
\cite[]{cylinderSens}.
In other words, the dynamical system has negative Lyapunov exponents.
In the state space, the steady fluid flow solution is a
(zero-dimensional) fixed point attractor.
For this flow field, the adjoint solution also quickly settles down to a
steady state when integrated backwards in time.

\begin{figure} \centering
\includegraphics[width=0.6\textwidth,trim=200 350 200 300,clip]
{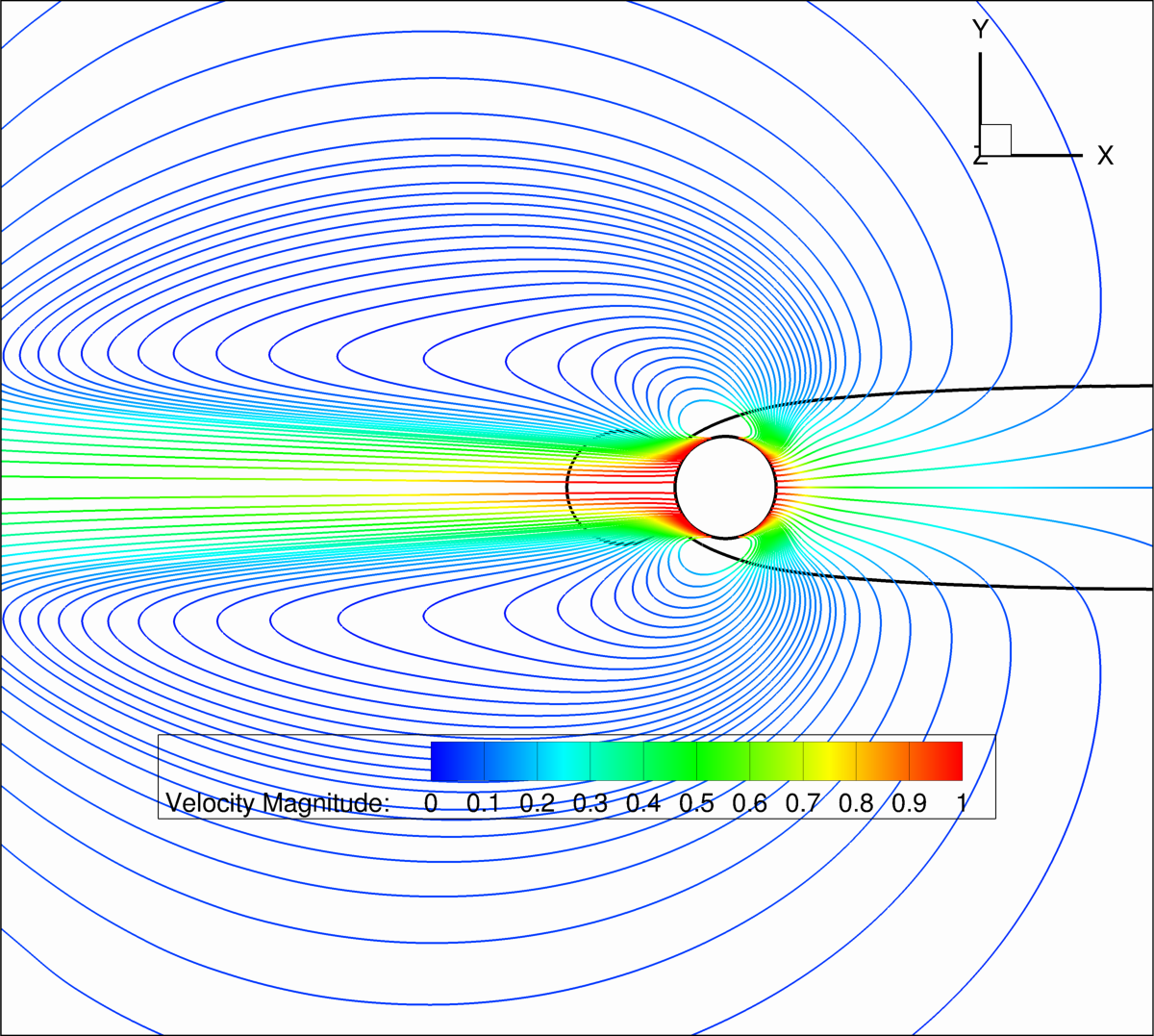}\hspace{0.3\textwidth}
\includegraphics[width=0.35\textwidth]{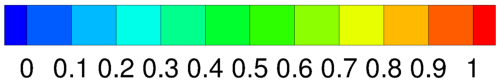}
\caption{The adjoint velocity field, $Re_D=20$.
The color indicates the magnitude of the adjoint field $\hat{\mathbf{u}}$.
The black lines are contours of streamwise flow velocity
at $0$ and $0.8$ freestream velocity.}
\label{f:ReD20}
\end{figure}

As shown in Figure \ref{f:ReD20}, the main feature of the adjoint field is
a jet of adjoint velocity, generated by the boundary condition
$\hat{\mathbf{u}}=(1,0,0)$ at the cylinder surface.  The jet propagates upstream
while being diffused rapidly by the high viscosity.
The adjoint boundary condition of $(1,0,0)$ also applies to the
downstream side of the cylinder; the main effect on the downstream is
similar to suction.  The apparently odd features of the upstream jet and
downstream suction is due to the fact that the adjoint evolves
backwards in time; thus the directionality of advection is reversed.
The jet and suction on the cylinder surface also
produce two large eddies on the upper and lower sides of the cylinder.

In this adjoint field, the net adjoint energy production is calculated
as $\iiint \hat{P}^{\hat{\mathcal{E}}}\,dV = 0.71$, and the net energy
dissipation $\iiint \hat{D}^{\hat{\mathcal{E}}}\,dV = 1.07$.
Because the adjoint field is steady state, the left hand side of the
energy balance equation (\ref{energy}) is 0.  Therefore, the net loss
of energy in the interior of the domain can only be compensated by
flux through the boundary.  This energy could come from the jet 
emanating from the front side of the cylinder.  We also observe that
the magnitude of the adjoint velocity $\hat{\mathbf{u}}$ is almost always less
than 1, which is determined by the boundary condition.  This may be
explained by the dominance of the dissipation term over the production
term in the energy balance.

\subsection{Adjoint flow field of periodic wake at $Re_D=100$}

\begin{figure}\centering
  \subfloat[near field, $t=0$]{\label{f:100a}
  \includegraphics[width=0.45\textwidth,trim=0 200 0 200,clip]{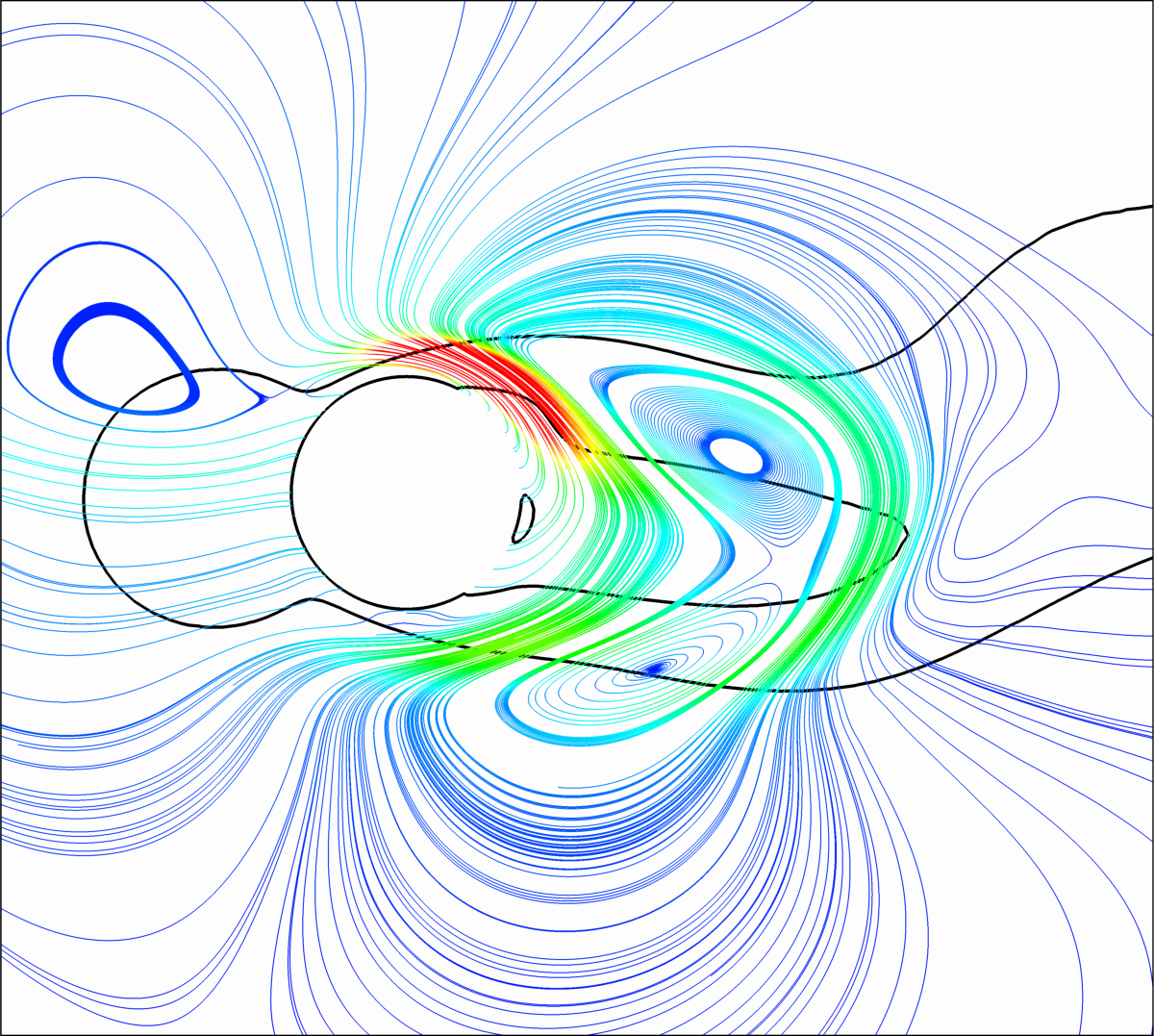}}
  \hspace{0.03\textwidth}
  \subfloat[far field, $t=0$]{\label{f:100b}
  \includegraphics[width=0.45\textwidth,trim=0 200 0 200,clip]{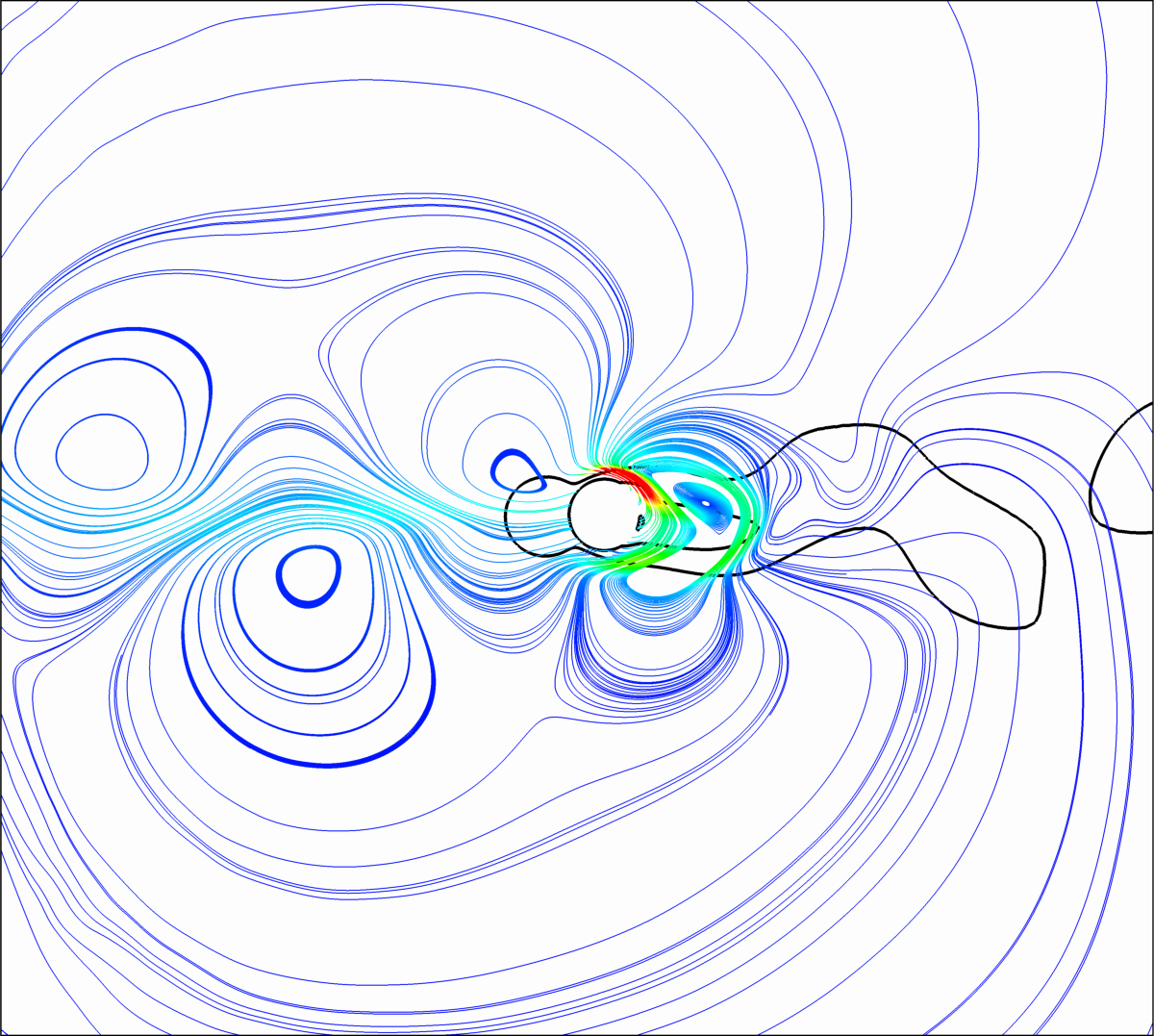}}
  \hspace{0.03\textwidth}
  \subfloat[near field, $t=1$]{\label{f:100c}
  \includegraphics[width=0.45\textwidth,trim=0 200 0 200,clip]{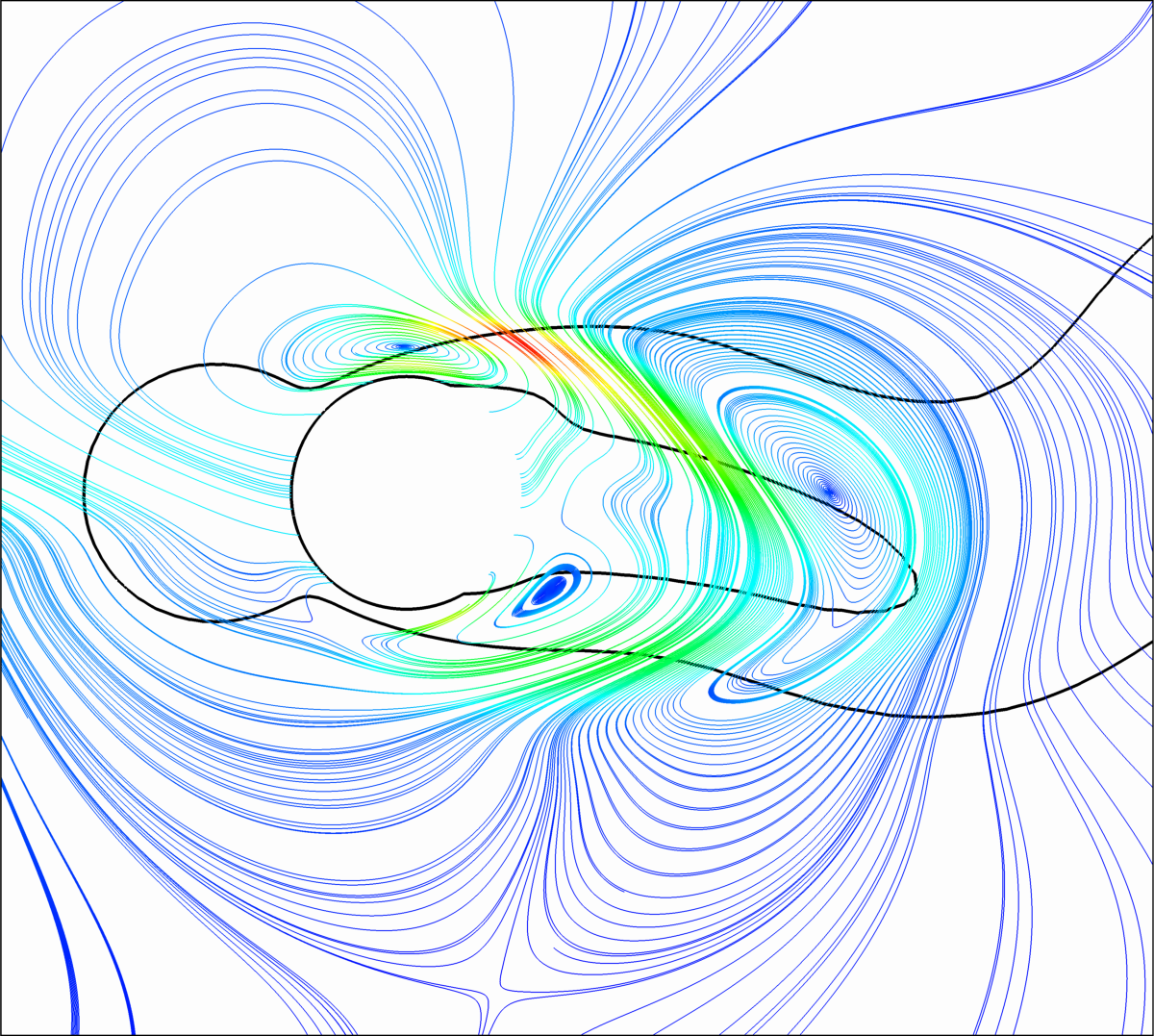}}
  \hspace{0.03\textwidth}
  \subfloat[far field, $t=1$]{\label{f:100d}
  \includegraphics[width=0.45\textwidth,trim=0 200 0 200,clip]{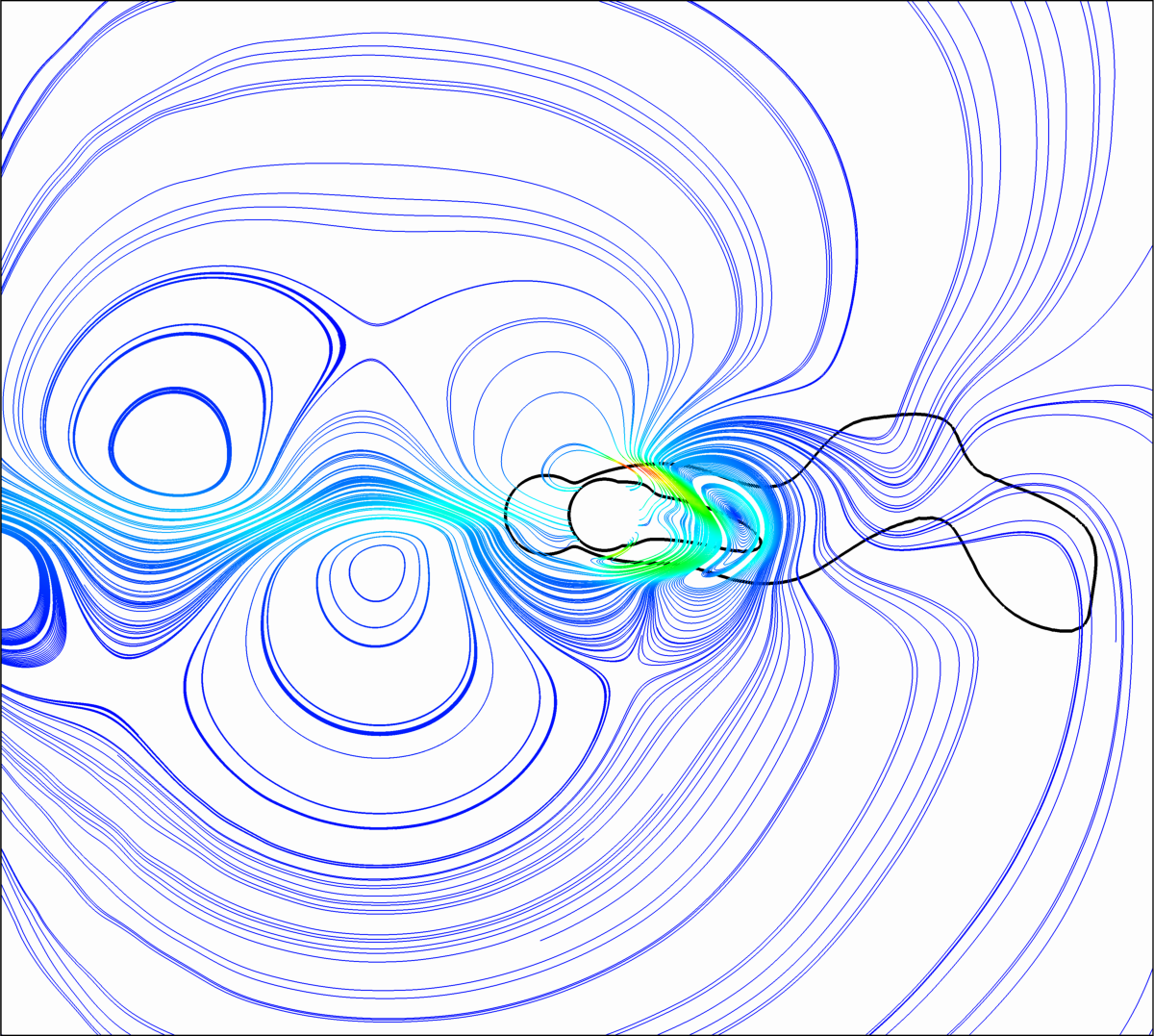}}
  \hspace{0.03\textwidth}
  \subfloat[near field, $t=2$]{\label{f:100e}
  \includegraphics[width=0.45\textwidth,trim=0 200 0 200,clip]{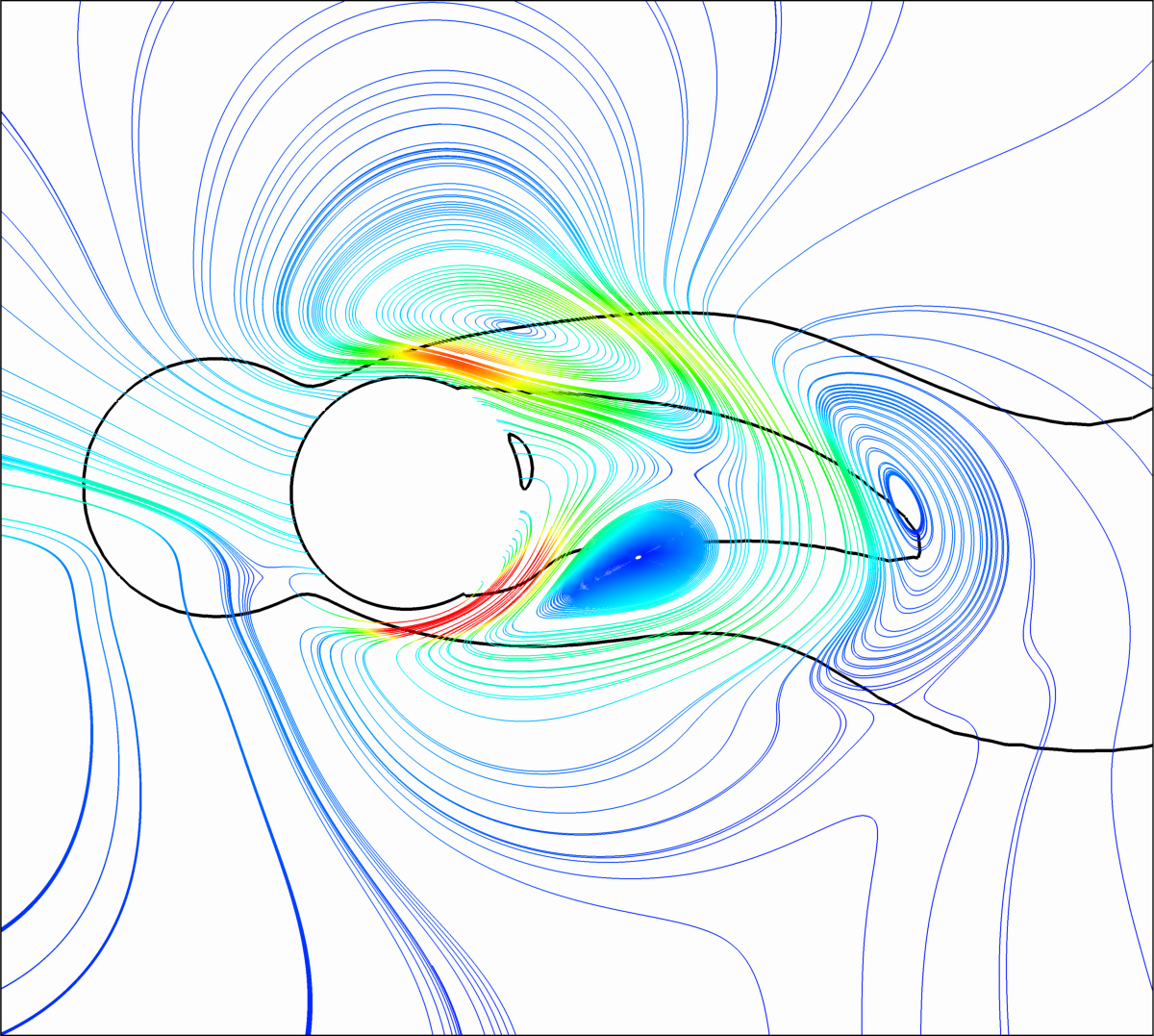}}
  \hspace{0.03\textwidth}                                   
  \subfloat[far field, $t=2$]{\label{f:100f}
  \includegraphics[width=0.45\textwidth,trim=0 200 0 200,clip]{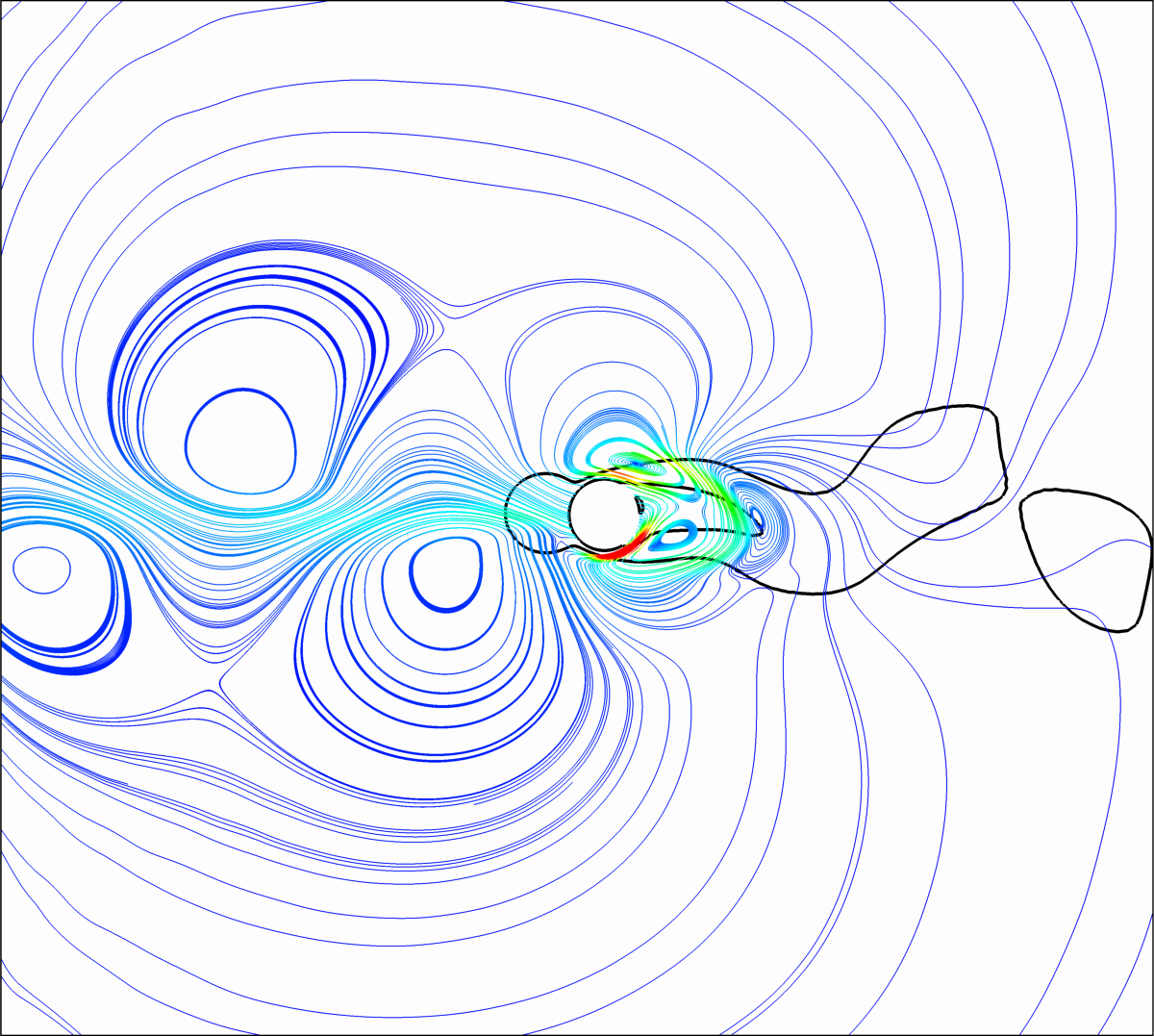}}
  \hspace{0.03\textwidth}                                   
  \subfloat{\includegraphics[width=0.4\textwidth]{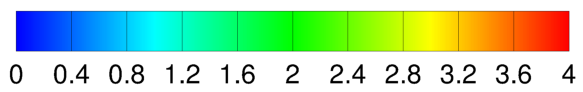}}
  \caption{The adjoint velocity field for about half of a period,
  $Re_D=100$.  The color indicates the magnitude of the adjoint field
  $\hat{\mathbf{u}}$.  The black lines are contours of the streamwise flow
  velocity at $0$ and $0.8$ freestream velocity.}
  \label{f:ReD100}
\end{figure}

\begin{figure} \centering
\includegraphics[width=0.8\textwidth,trim=35 0 0 0,clip]{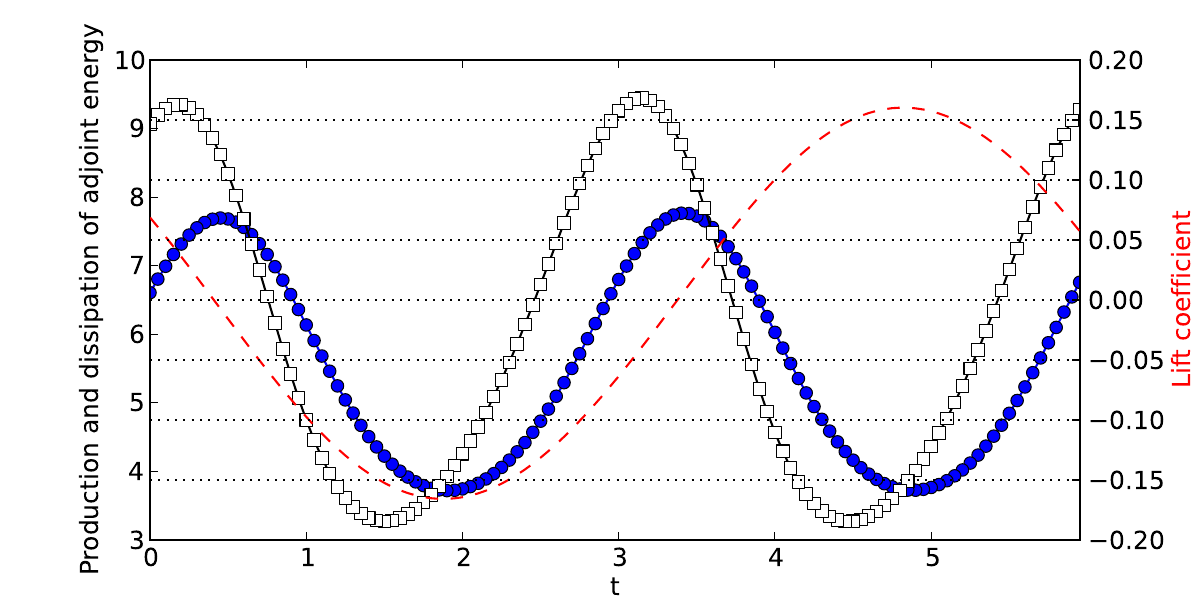}
\caption{The energy balance of the adjoint field at $Re_D=100$.  The
blue, filled circles represent the net energy production
$\iiint \hat{P}^{\hat{\mathcal{E}}}\,dV$; the black, open squares
represent the net energy dissipation
$\iiint \hat{D}^{\hat{\mathcal{E}}}\,dV$; the red, dashed line is
the lift coefficient of the cylinder, which serves to identify the phase
of the vortex shedding.}
\label{f:energy100}
\end{figure}

At Reynolds number $Re_D=100$, the fluid flow exhibits periodic von
Karman vortex shedding.  Through Floquet analysis
\cite[]{cylinderSens,CambridgeJournals:379198}, it can be shown that
any small perturbation to the periodic fluid flow will asymptotically decay
to the periodic state; however, there can be a non-decaying phase difference
between the vortex shedding of the unperturbed fluid flow and the perturbed
fluid flow, even after the transient disturbance settles down.
In other words, the system has a Lyapunov exponent that is exactly equal to 0,
and all other Lyapunov exponents are negative.
This is a common feature in dynamical systems with a limit cycle
attractors.  In the state space, the attractor is a one dimensional
closed curve.  The points on the curve represent time snapshots of the fluid
flow.  The adjoint field of the periodic flow also settles down to a
periodic state in about 2 to 3 shedding periods.

Figure \ref{f:ReD100} illustrates the evolution of the adjoint field
via the stream trace of the adjoint velocity, colored by the adjoint
velocity magnitude.  The streamwise fluid velocity contours (black) indicate
the phase of the vortex shedding and the location of the shear layers.
Because the vortex shedding is symmetric, we show
three time snapshots covering about half of a period, and
the other half of the period is symmetric to the half-period shown.
The adjoint field evolves backwards in time, therefore,
a good way of examining the dynamics over an entire period is looking at
Figures \ref{f:100e}, then \ref{f:100c} and \ref{f:100a}, and then the
upside-down versions of Figure \ref{f:100e}, \ref{f:100c},
\ref{f:100a}, in that order.
The right column in Figure (\ref{f:ReD100}) shows the zoomed-out views
of the same adjoint fields.

In Figure \ref{f:ReD100}, the most significant feature in the
adjoint field is the bean-shaped eddies behind the
cylinder.  When viewed backwards in time, these bean-shaped patterns
first form at about 2 diameters downstream of the cylinder
(Figure \ref{f:100e}),
and grow larger and stronger as they propagate upstream in the wake
(Figures \ref{f:100c} and \ref{f:100a}).  The mechanism responsible for
generation and amplification of these bean-shaped eddy structures
is described in Section \ref{s:circulation}, and qualitatively
depicted in Figure \ref{wakeadj}.

The bean-shaped adjoint eddy breaks up into two elongated eddies
at about 1 diameter downstream of the cylinder (Figure \ref{f:100e}).
Both eddies are aligned in oblique directions to the shear layer.
As both eddies approach the cylinder surface,
they experience strong dissipation (Figure \ref{f:100c}).  One of the
two eddies (the lower one in Figure \ref{f:100c}) is trapped in the wake and
disappears.  The other eddy (the upper one in
Figure \ref{f:100c}) propagates upstream of the cylinder in
a relatively uniform fluid flow
(the upper left eddy in Figure \ref{f:100a}).  As these eddies
propagate further upstream, they spread out and become more circular by the
force of viscosity (Figures \ref{f:100f}, \ref{f:100d} and \ref{f:100b}).

The jet of adjoint velocity in front of the cylinder is still apparent
in the adjoint field.  It has a magnitude smaller than the
adjoint circulation in the wake, and meanders unsteadily between the
vortices propagating upstream.

The energy balance of the periodic adjoint field is shown in Figure
\ref{f:energy100}.   The net production and dissipation in the interior
of the domain have similar magnitudes, and overtake each other twice
per period.  Averaged over time, the dissipation is slightly more than
the self-production of adjoint energy in the interior.  This net loss of
energy could be compensated by the jet emanating from the cylinder
surface.

\subsection{Adjoint flow field of turbulent wake at $Re_D=500$}

\begin{figure}\centering
  \subfloat[$t=2$]
  {\includegraphics[width=0.48\textwidth,trim=200 300 30 300,clip]
  {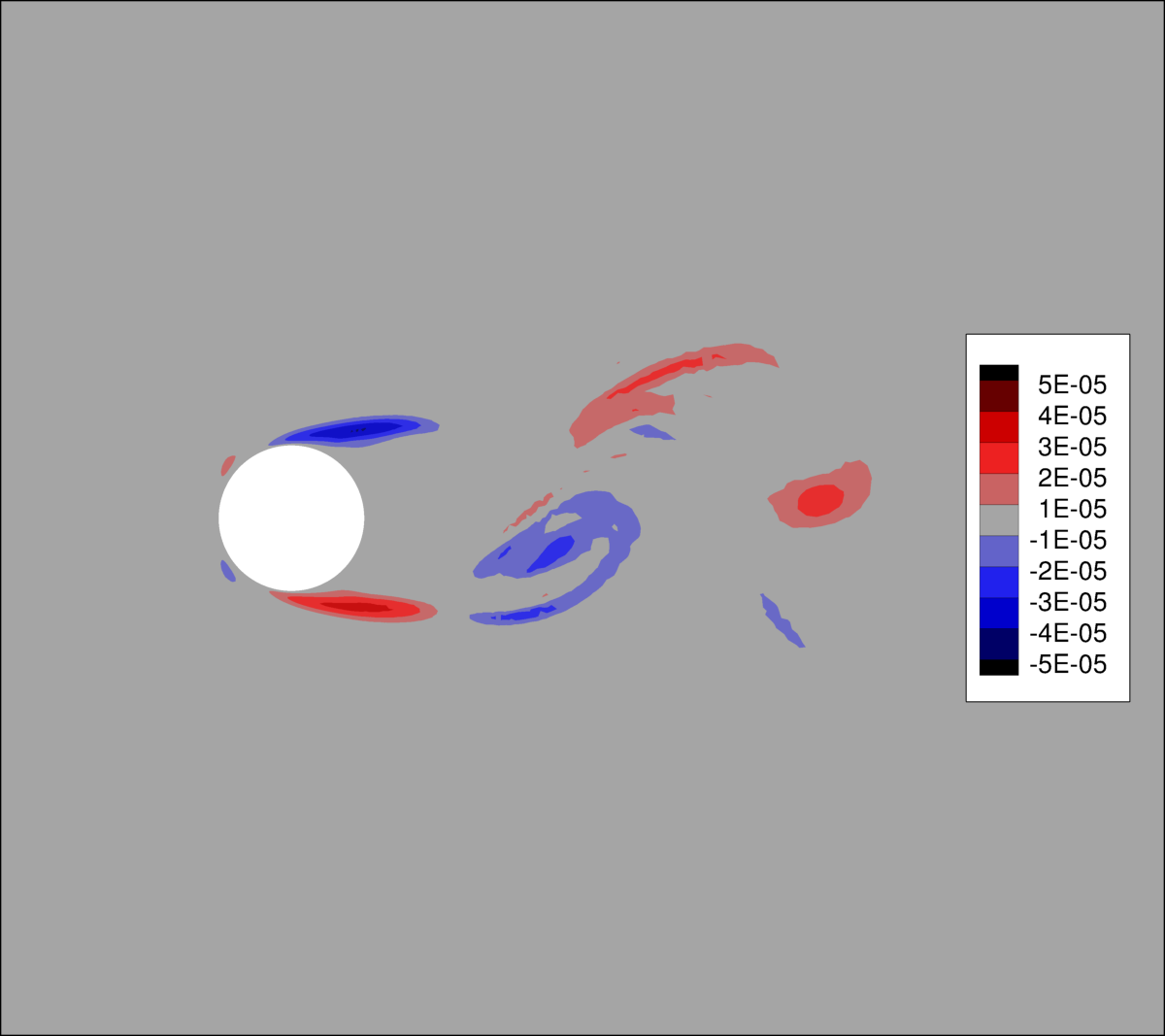}}
  \hspace{0.01\textwidth}
  \subfloat[$t=22$]
  {\includegraphics[width=0.48\textwidth,trim=200 300 30 300,clip]
  {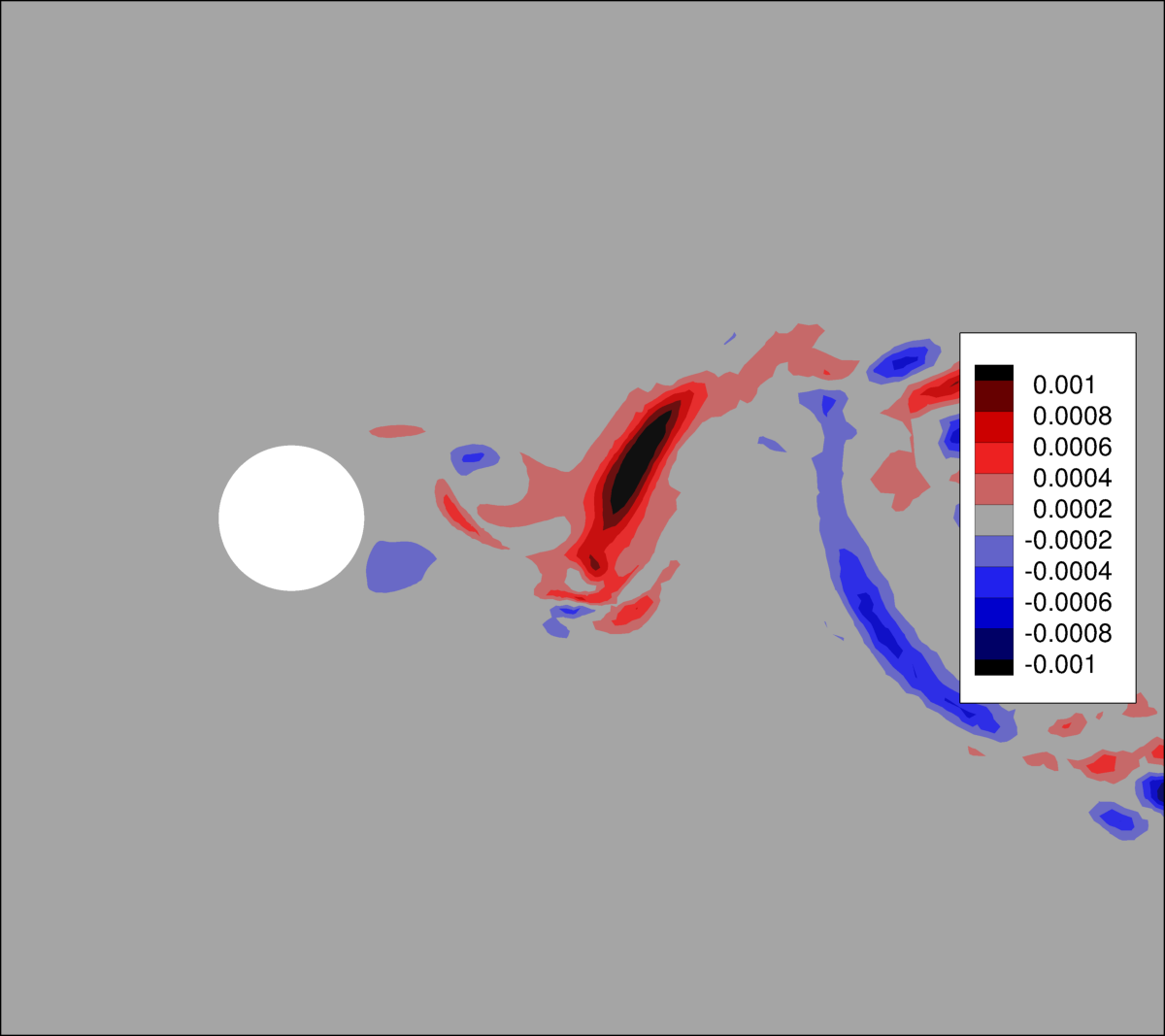}}
  \hspace{0.01\textwidth}
  \subfloat[$t=42$]
  {\includegraphics[width=0.48\textwidth,trim=200 300 30 300,clip]
  {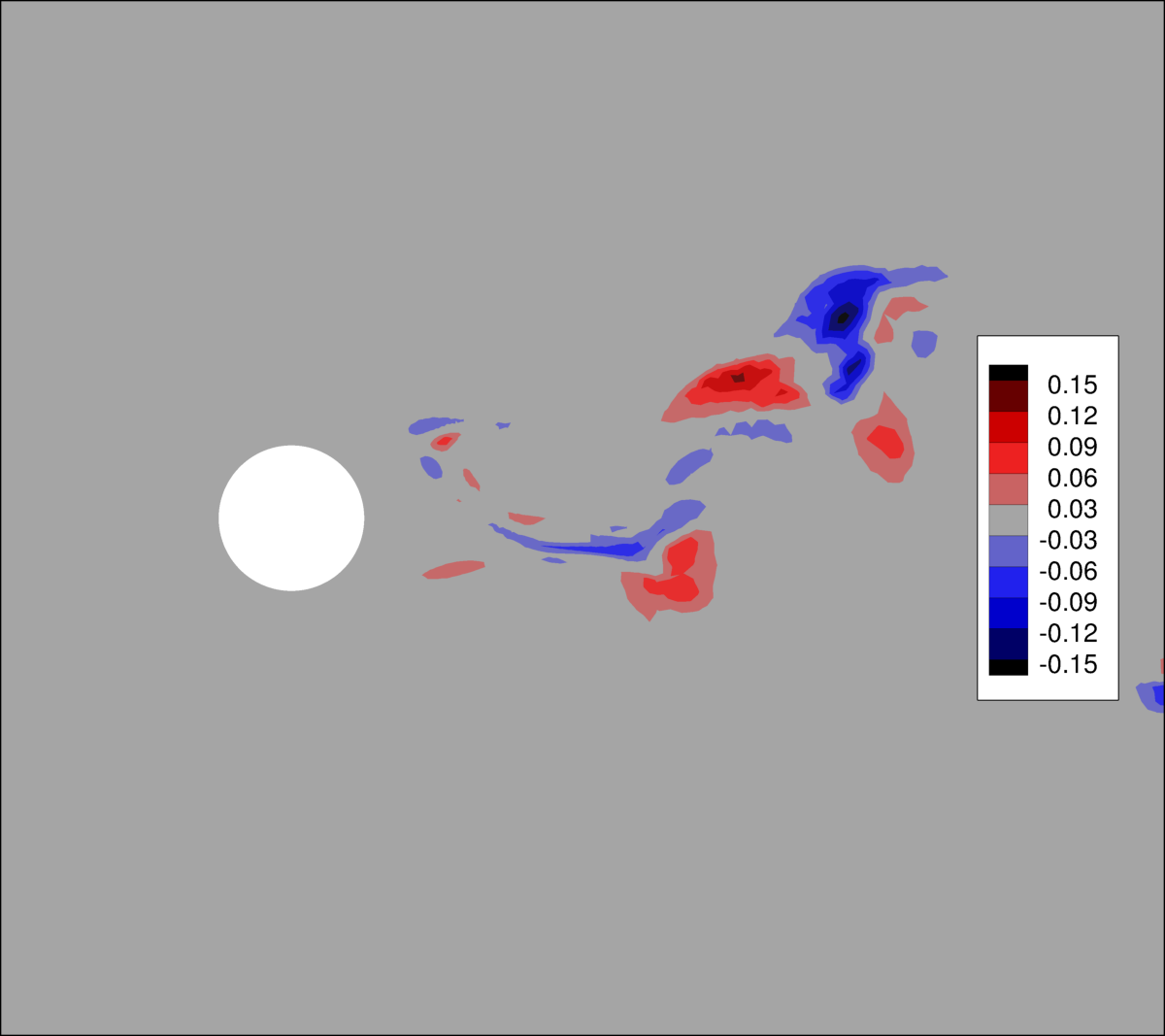}}
  \hspace{0.01\textwidth}
  \subfloat[$t=62$]
  {\includegraphics[width=0.48\textwidth,trim=200 300 30 300,clip]
  {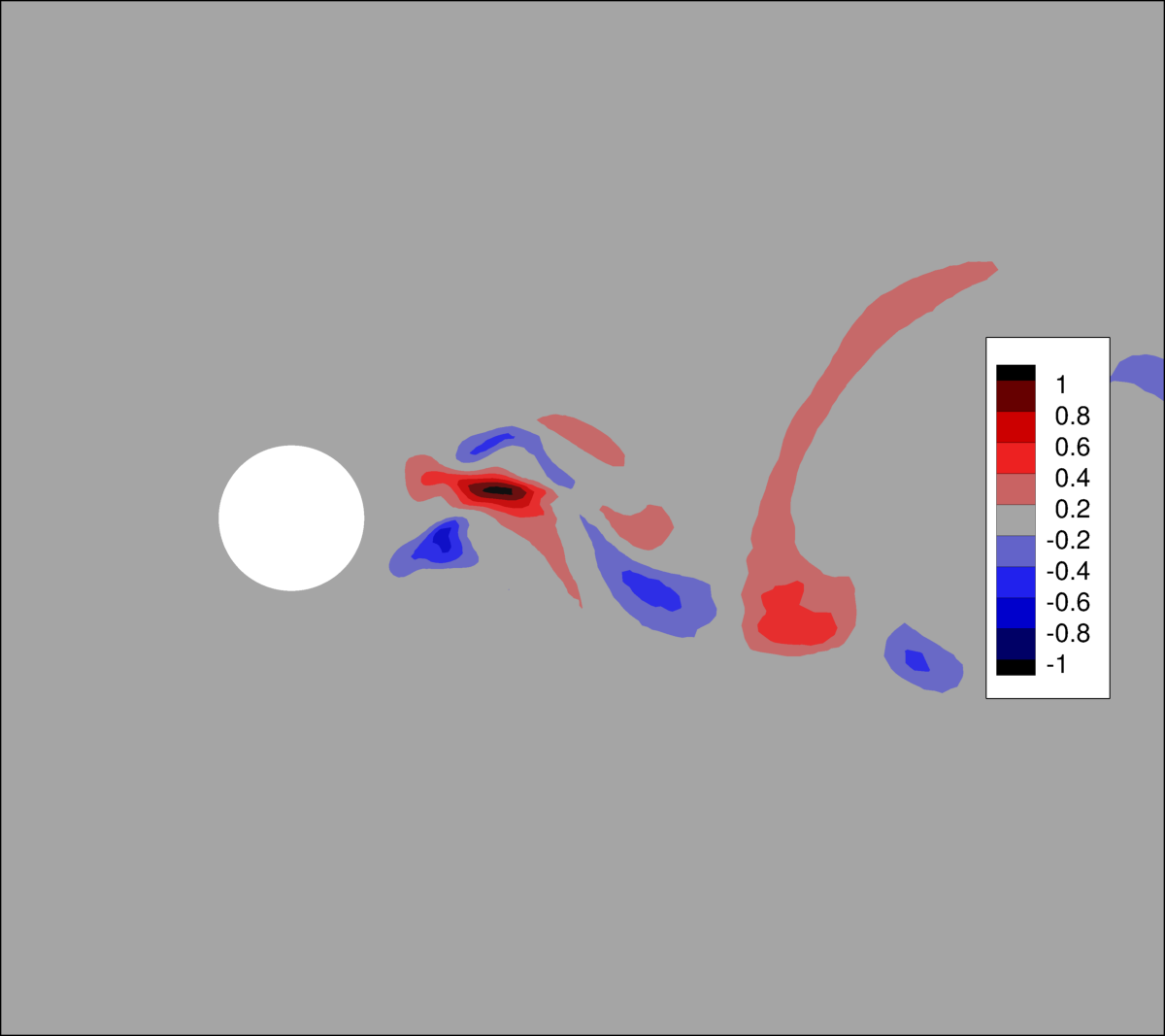}}
  \caption{The ``butterfly effect'' of flow at $Re_D=500$.
  The plots show the difference in spanwise velocity between a perturbed
  flow field and the unperturbed flow field, at different time snapshots.}
\label{f:butter}
\end{figure}

At Reynolds number $Re_D\approx 189$, a series of complex transitions
to turbulence starts.  Accurately capturing the unsteady flow field
during the transition may require a very large spanwise domain
\cite[]{CambridgeJournals:379198}.
However, at Reynolds number $Re_D=500$, \cite{CambridgeJournals:398060}
showed that the wake exhibits fully turbulent behavior even with a
modest spanwise extent of $\pi$ diameters.  In this paper, a spanwise
extent of $4$ diameters is used.

In the chaotic, turbulent wake structure, an infinitesimal
perturbation could grow exponentially until saturated by nonlinearity.
This phenomenon is known as the ``butterfly effect''.
Figure \ref{f:butter} demonstrates this butterfly effect in the
$Re_D=500$ case.  The four plots show the difference between the spanwise
velocity of a perturbed flow field and that of an unperturbed flow
field.  The unperturbed flow field has been time-integrated to
statistical equilibrium at $t=0$; the perturbed flow field is a copy of
the unperturbed flow field at $t=0$, plus a perturbation of magnitude
$10^{-5}$ at one diameter downstream of the cylinder.  Starting from
their slightly different initial conditions at $t=0$, the perturbed and
unperturbed flow fields are then time-integrated forwards independently,
with exactly the same boundary conditions.  Figure \ref{f:butter} shows
that the difference between the two flow fields, which starts at
$10^{-5}$ at $t=0$, grows larger as time progresses.  At $t=62$, the
difference becomes order 1, as large as the freestream velocity.
At that point, the two flow fields are significantly different.

The divergence of a small perturbation is a signature of chaotic
dynamics.  It implies that almost any quantity that depends on the
flow field at a later time, e.g. $t=62$, is very sensitive to almost
any small perturbation made at an earlier time, e.g. $t=0$.
This has significant implications for the adjoint field, which contains
information about sensitivities: the adjoint field at time $t$
indicates sensitivity of an overall quantity of interest to potential
perturbations made at $t$.  This sensitivity can be very large if
the quantity of interest depends on the flow solution at a time much later
than $t$, as implied by the butterfly effect of chaos.
In our calculation, the quantity of interest embedded in the adjoint
equation is an integral over a fixed range of time $[0,T]$;
therefore, we expect the adjoint field at time $t$ to be very much larger for
$t\ll T$.

\begin{figure} \centering
\includegraphics[width=0.83\textwidth,trim=0 0 0 0,clip]{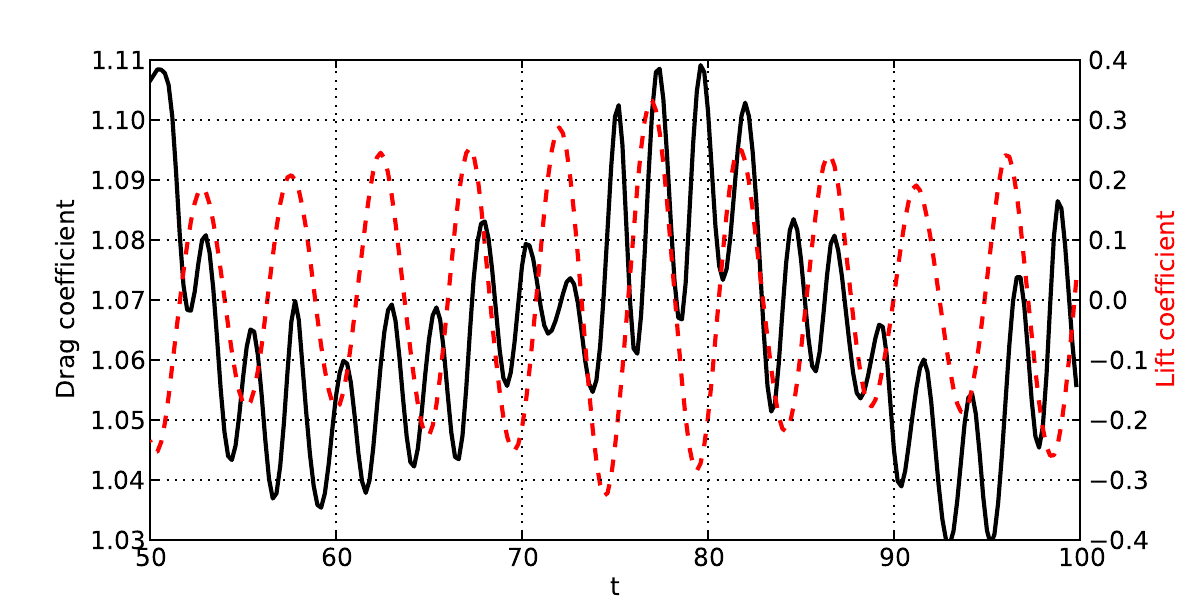}
\includegraphics[width=0.83\textwidth,trim=0 0 0 0,clip]{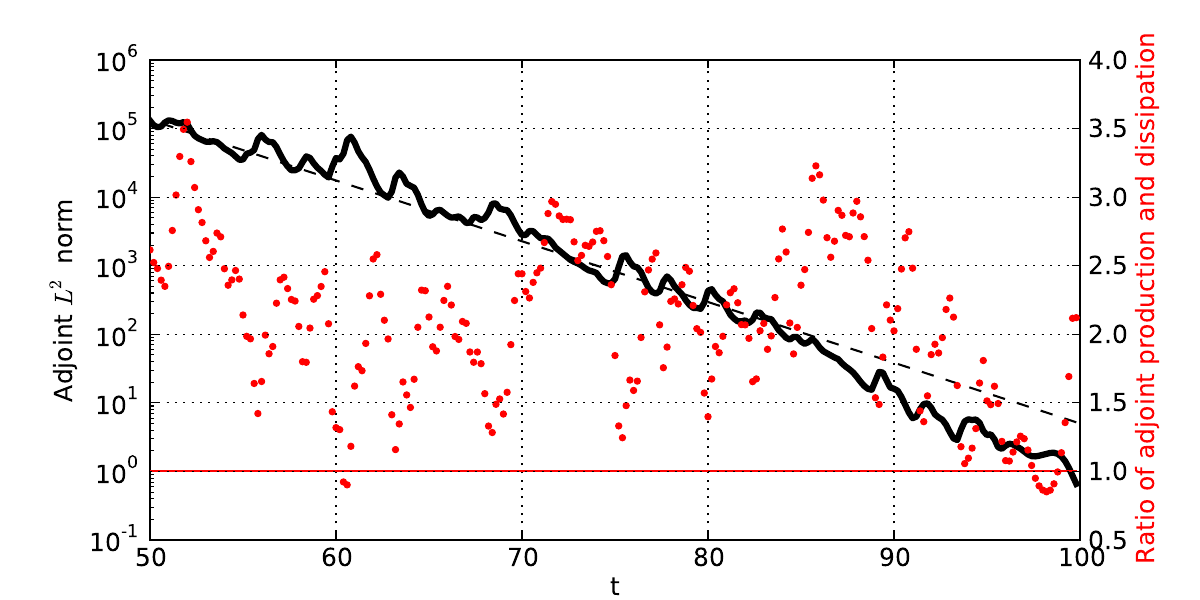}
\caption{The upper plot shows the lift and drag coefficients as
functions of time at $Re_D=500$.  The lower plot shows the $L^2$ norm
of the adjoint solution (solid black line), compared to an estimate
based on the Lyapunov exponent (dashed line).  
The ratio of net adjoint energy production and dissipation
$\iiint \hat{P}^{\hat{\mathcal{E}}}\,dV / \iiint
 \hat{D}^{\hat{\mathcal{E}}}\,dV$ is plotted as red dots.}
\label{f:energy500}
\end{figure}

It is indeed observed that the adjoint field has a larger magnitude at
smaller $t$.  As shown in the lower plot of Figure \ref{f:energy500},
the magnitude of the adjoint equation increases
exponentially as $t$ decreases.  This indicates that the quantity of
interest is exponentially more sensitive to perturbations at earlier times.
This agrees with our observation in Figure \ref{f:butter} that the flow
fields at later times are exponentially more sensitive to a perturbation
at $t=0$.  In addition, the rate at which the adjoint field increases
agrees with the Lyapunov exponent, as indicated by the dashed line in
Figure \ref{f:energy500}; the Lyapunov exponent is estimated from the
rate at which the perturbed and unperturbed fields diverge
(as shown in Figure \ref{f:butter}).

The energy balance of the chaotic adjoint field is also shown in Figure
\ref{f:energy500}.   The net production in the interior
of the domain almost always exceeds the net dissipation, as indicated
by the production dissipation ratio.  The adjoint energy production in
the interior is the main contributor of the exponential growth of the
adjoint solution as time goes backwards.  This is because the interior energy
production term $\hat{P}^{\hat{\mathcal{E}}}$ in Equation (\ref{energy})
is proportional to the squared magnitude of the adjoint solution
$\hat{\mathbf{u}}$, while the contribution of energy from the boundary is only
proportional to the magnitude of $\hat{\mathbf{u}}$.   As the adjoint solution
grows larger, the boundary condition becomes less important compared to
the interior dynamics.

\begin{figure}\centering
  \includegraphics[width=0.47\textwidth,trim=0 400 0 0,clip]{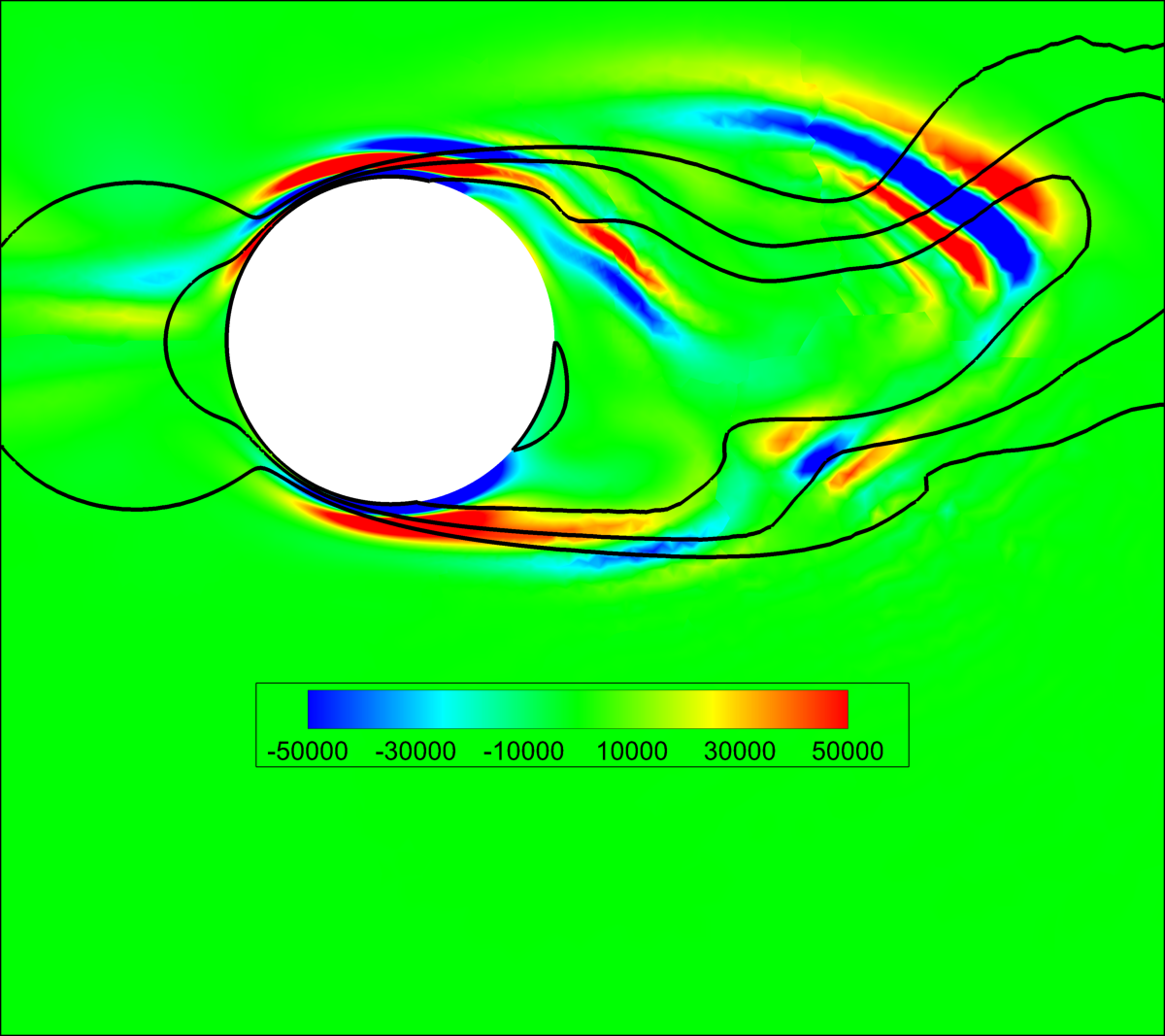}
  \hspace{0.01\textwidth}
  \includegraphics[width=0.47\textwidth,trim=0 400 0 0,clip]{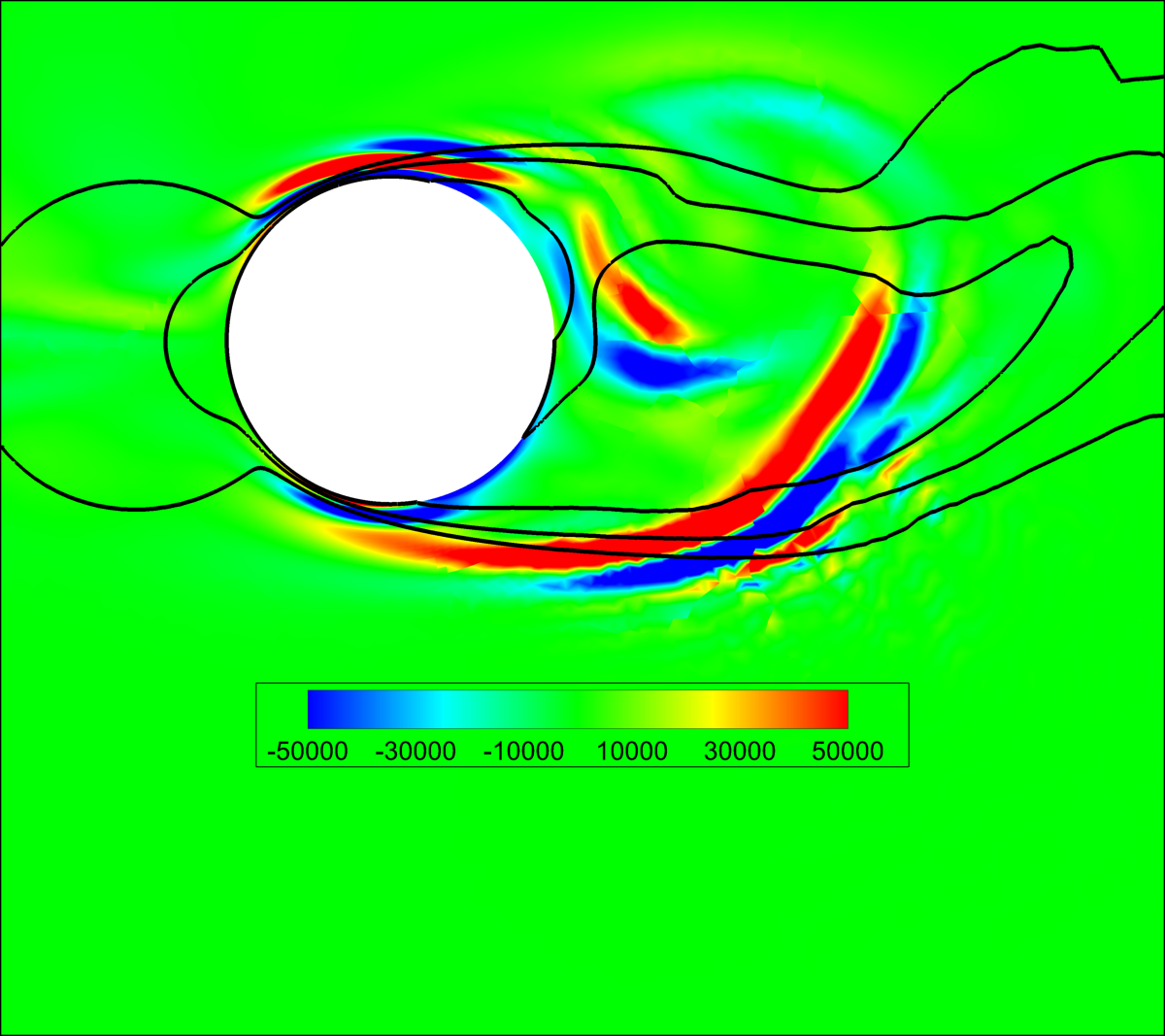}\\
  \vspace{0.02\textwidth}
  \includegraphics[width=0.47\textwidth,trim=0 250 0 0,clip]{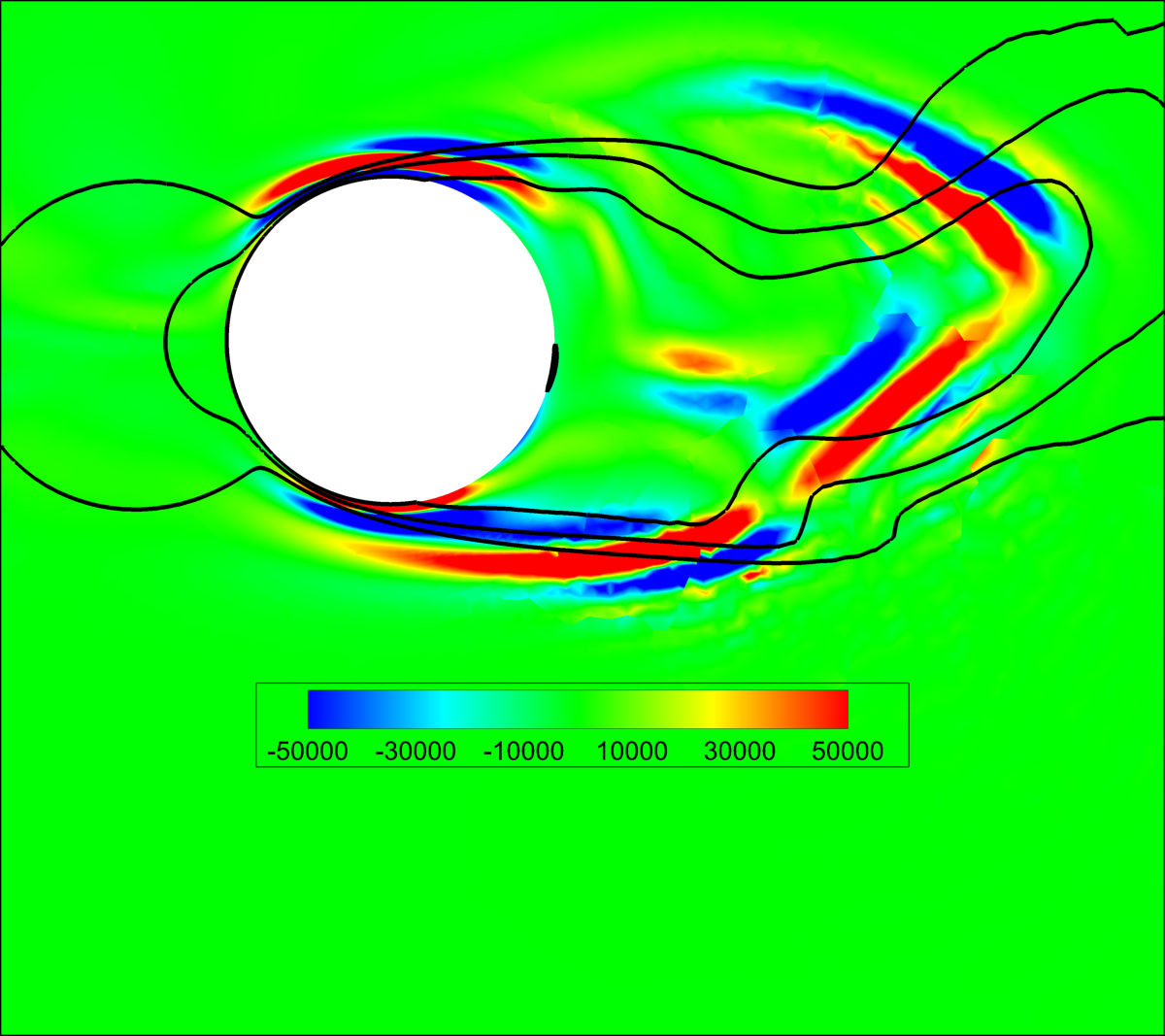}
  \hspace{0.01\textwidth}                               
  \includegraphics[width=0.47\textwidth,trim=0 250 0 0,clip]{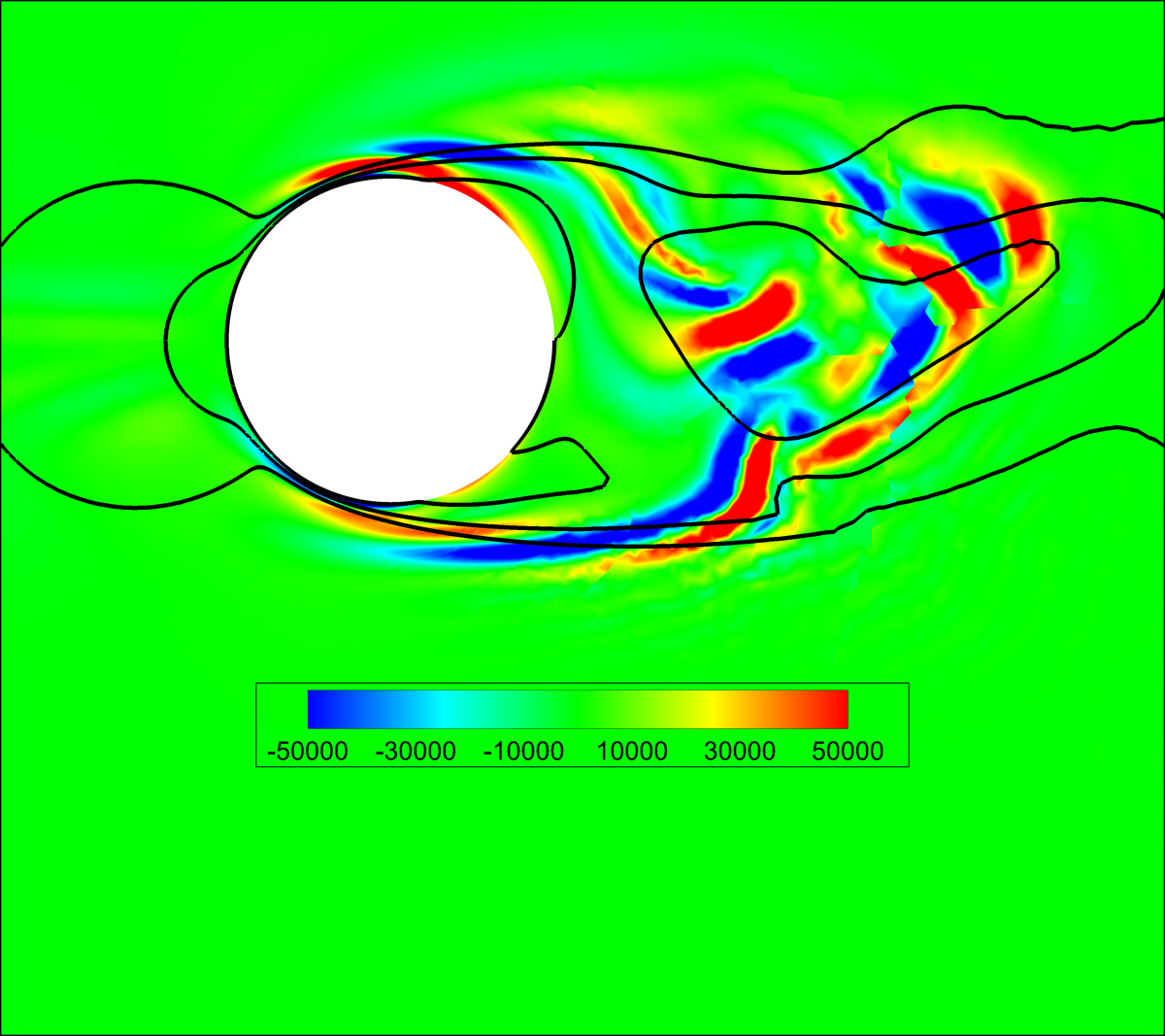}
  \caption{The spanwise component of the adjoint vorticity
  $\nabla\times\hat{\mathbf{u}}$ at $Re_D=500$.  The four plots show the spanwise
  vorticity at different spanwise sections at time $t=T_1-38$.
  The black lines are contours of streamwise flow velocity
  at $0$, $0.4$ and $0.8$ freestream velocity.}
\label{f:ReD500}
\end{figure}

Figure \ref{f:ReD500} visualizes the diverging adjoint field
by plotting the spanwise adjoint vorticity $\nabla\times\hat{\mathbf{u}}$
at four different spanwise sections at $t=T_1-38$, i.e., after 38
time units of adjoint time integration.  The field concentrates almost
entirely in the near-wake region.  The adjoint eddies
appear in mostly parallel streaks that are oblique to the shear layers
in the wake.  These structures are consistent with our analysis in
Section \ref{s:circulation}, as sketched in Figure \ref{adjointcirc}.

The jet of adjoint velocity in front of the cylinder, a
significant feature in the lower Reynolds number cases, is no longer
visible.  This can be explained by comparing the adjoint energy production in
the interior $\hat{P}^{\hat{\mathcal{E}}}$ (Equation (\ref{energy}))
and adjoint energy production on the boundary
$\hat{P}^{BC}$ (Equation (\ref{eneprodbc})).
As the magnitude of the adjoint field becomes larger, the contribution
of energy from the boundary condition increases linearly, while the
contribution of energy from the interior production term
$\hat{P}^{\hat{\mathcal{E}}}$ increases quadratically.  Because the
energy of the jet comes from the boundary condition, its magnitude
is relatively small when the adjoint field becomes very large.

\section{Conclusions} 
\label{s:conclusion}

The solution of the adjoint equation with drag being the quantity of
interest can
be viewed as a nondimensional transfer function of small momentum
perturbations.  We analyze the solution of the adjoint equation with
conventional fluid mechanical methods, including studying the kinetic
energy of the adjoint field and the evolution of circulation along a
closed material contour.  The result of our analysis shows that the
adjoint kinetic energy is not conserved; in particular, the adjoint field can
exponentially amplify along certain eigen-directions of the flow shear
rate tensor $\frac12 (\nabla \mathbf{u} + \nabla \mathbf{u}^T)$.
Analysis of circulation dynamics reveals
that the dynamics of the adjoint equation should preferentially amplify
elongated eddies that are aligned with converging directions of
a shear flow.  In a parallel shear flow, the amplified
eddies should be elongated along a 45 degrees angle direction relative
to the shear layer.

Numerical solutions of the adjoint equation are performed for a circular
cylinder at Reynolds numbers $Re_D=20, 100$ and $500$.  At $Re_D=20$,
both the flow solution and the adjoint solution are steady.  
Downstream of the cylinder, the adjoint field has streamlines similar
to that of fluid being sucked into the downstream part of the cylinder;
upstream of the cylinder, the adjoint field has streamlines similar
to that of fluid being ejected towards upstream of the cylinder,
forming a jet-like structure.
These features in the adjoint field are consistent with the physical
interpretation of the adjoint field.
At $Re_D = 100$, both the flow field and the adjoint field are 2D and
periodic.  We observe bean-shaped eddies in the adjoint field forming in
the near-wake region of the cylinder.  These eddies form because they
are preferentially amplified by the dynamics of the adjoint equation in
shear flow.  At $Re_D=500$, the flow in the wake is turbulent;
the entire adjoint field is amplified exponentially as time evolves
backwards.  This amplification is caused by the production of adjoint
energy in the interior dominating viscous dissipation.  The rate
of exponential amplification is consistent with the Lyapunov exponent of
the turbulent flow.  The dominant structure in the adjoint field is
thin, elongated eddies in the near-wake region, created by preferential
amplification of these eddies in the shear layers of the near wake.
 
\begin{acknowledgments}
We would like to acknowledge financial support from the subcontract of
DOE’s Stanford PSAAP to MIT, AFOSR support under STTR contract
FA9550-12-C-0065 under Dr. Fariba Fahroo, and support from the NASA
Fundamental Aeronautics Program under Dr. Harold Atkins.
\end{acknowledgments}

\bibliographystyle{jfm}

\bibliography{master}

\appendix
\section{Derivation of the adjoint equation}
\label{s:app}

The derivation starts with the linearized Navier-Stokes equation
(\ref{linearized}) that describes the small change in the flow field
$\delta \mathbf{u}$ caused by an infinitesimal body force
$\delta\mathbf{f}$ in the interior of the flow field.
By taking the inner product of the adjoint field $\hat{\mathbf{u}}(x,t)$ with the
linearized momentum equation in Equation (\ref{linearized}), and multiplying
the adjoint pressure $\hat{p}$ with the divergence-free condition in Equation
(\ref{linearized}), we obtain
\begin{equation} \label{aL} \begin{split}
 & \underbrace{\rho \frac{\partial \delta \mathbf{u}}{\partial t} \cdot
 \hat{\mathbf{u}}}_{(L1)}
 + \underbrace{\rho\, \mathbf{u}\cdot \nabla \delta \mathbf{u} \,\cdot\hat{\mathbf{u}}}_{(L2)}
 + \underbrace{\rho\, \delta \mathbf{u}\cdot \nabla \mathbf{u} \,\cdot\hat{\mathbf{u}}}_{(L3)}
 + \underbrace{\nabla \delta p \,\cdot\hat{\mathbf{u}}}_{(L4)}
 - \underbrace{\mu \nabla \cdot \nabla \delta \mathbf{u} \,\cdot\hat{\mathbf{u}}}_{(L5)}
 = \delta\mathbf{f} \,\cdot\hat{\mathbf{u}} \\
 & \underbrace{\hat{p}\;\nabla\cdot \delta \mathbf{u}}_{(L6)} = 0
\end{split} \end{equation}

The adjoint equation (\ref{adjoint}) is designed to be
``symmetric'' to the linearized Navier-Stokes equation.
By taking the inner product of the velocity field perturbation $\delta u(x,t)$
with the adjoint momentum equation in Equation (\ref{adjoint}), and multiplying
the pressure field perturbation $\delta p$ with the divergence-free condition
in Equation (\ref{adjoint}), we obtain
\begin{equation}  \label{aA} \begin{split}
 & \underbrace{\rho \frac{\partial \hat{\mathbf{u}}}{\partial t} \cdot
 \delta \mathbf{u}}_{(A1)}
 + \underbrace{\rho\, \mathbf{u}\cdot \nabla \hat{\mathbf{u}} \,\cdot\delta \mathbf{u}}_{(A2)}
 - \underbrace{\rho\, \delta \mathbf{u}\cdot \nabla \mathbf{u} \,\cdot\hat{\mathbf{u}}}_{(A3)\equiv(L3)}
 + \underbrace{\nabla \hat{p} \,\cdot\delta \mathbf{u}}_{(A6)}
 + \underbrace{\mu \nabla \cdot \nabla \hat{\mathbf{u}} \,\cdot\delta \mathbf{u}}_{(A5)}
 = 0 \\
 & \underbrace{\delta p\;\nabla\cdot \hat{\mathbf{u}}}_{(A4)} = 0
\end{split} \end{equation}

By adding corresponding terms together and integrating over either
space or time, we obtain
\begin{equation}
\int_0^T (L1) + (A1) \,dt =
  \rho\, \delta \mathbf{u}\cdot \hat{\mathbf{u}} \bigg|_0^T = 0
\end{equation}
because $\delta \mathbf{u}=0$ at $t=0$ and $\hat{\mathbf{u}}=0$ at $t=T$:
\begin{equation} \begin{split}
\iiint ((L2) + (A2)) \,dV &= \iiint \rho\, \mathbf{u}\cdot \nabla
  (\delta \mathbf{u} \cdot \hat{\mathbf{u}})\, dV = \oiint \rho\, (\mathbf{u} \cdot \mathbf{n})
  (\delta \mathbf{u} \cdot \hat{\mathbf{u}})\, ds \\
\iiint ((L4) + (A4)) \,dV &= \iiint \nabla\cdot (\delta p\,\hat{\mathbf{u}})
  \, dV = \oiint \delta p\, (\hat{\mathbf{u}} \cdot \mathbf{n}) \, ds \\
\iiint ((L6) + (A6)) \,dV &= \iiint \nabla\cdot (\hat{p}\,\delta \mathbf{u})
  \, dV = \oiint \hat{p}\, (\delta \mathbf{u} \cdot \mathbf{n}) \, ds \\
\iiint ((L5) + (A5)) \,dV &=
  \oiint \mu (\mathbf{n} \cdot \nabla \hat{\mathbf{u}} \,\cdot\delta \mathbf{u}
                           - \mathbf{n} \cdot \nabla \delta \mathbf{u} \,\cdot\hat{\mathbf{u}})\,ds
\end{split} \end{equation}
Therefore, by adding equations (\ref{aL}) and (\ref{aA}), we obtain
\begin{equation} \label{alast}\begin{split}
& \int_0^T \iiint \delta\mathbf{f} \,\cdot\hat{\mathbf{u}}\,dV\,dt \\
=& \int_0^T
\oiint \Big(\rho\, (\mathbf{u} \cdot \mathbf{n}) (\delta \mathbf{u} \cdot \hat{\mathbf{u}})
+ \hat{\mathbf{u}} \cdot (\delta p\, \mathbf{n} - \mu\, \mathbf{n} \cdot \nabla \delta \mathbf{u})
+ \delta \mathbf{u} \cdot (\hat{p}\, \mathbf{n} + \mu\, \mathbf{n} \cdot
\nabla \hat{\mathbf{u}}) \Big)
\,ds \;dt
\end{split} \end{equation}
For external flow problems, we let the linearized and adjoint
Navier-Stokes equations satisfy the boundary conditions
\begin{equation}\begin{aligned}
& \delta{\mathbf{u}} = (0,0,0)\;, & & \hat{\mathbf{u}} = (1,0,0)\;, & & \mbox{at the walls}\\
& \delta{\mathbf{u}} = (0,0,0)\;, & & \hat{\mathbf{u}} = (0,0,0)\;, & & \mbox{at the far field}
\end{aligned} \end{equation}
For internal flow problems, we enforce a different set of boundary conditions
\begin{equation}\begin{aligned}
& \delta{\mathbf{u}} = (0,0,0)\;, & & \hat{\mathbf{u}} = (1,0,0)\;, & & \mbox{at the walls}\\
&\rho\, (\mathbf{u} \cdot \mathbf{n}) \delta \mathbf{u}
+ \delta p\, \mathbf{n} - \mu\, \mathbf{n} \cdot \nabla \delta \mathbf{u} = 0\;, &&
\mu\,\mathbf{n} \cdot \nabla \hat{\mathbf{u}} + \hat{p}\, \mathbf{n} = 0\;,&& \mbox{at the inlets}\\
&\rho\, (\mathbf{u} \cdot \mathbf{n}) \hat{\mathbf{u}}
+ \hat{p}\, \mathbf{n} + \mu\,\mathbf{n} \cdot \nabla \hat{\mathbf{u}} = 0 \;,&&
\mu\, \mathbf{n} \cdot \nabla \delta \mathbf{u} - \delta p\, \mathbf{n} = 0\;, && \mbox{at the outlets}
\end{aligned} \end{equation}
Either set of boundary conditions can be combined with Equation (\ref{alast})
to obtain
\begin{equation}
 \int_0^T \iiint \delta\mathbf{f} \,\cdot\hat{\mathbf{u}}\,dV\,dt \\
= \int_0^T \int_{S} 
\hat{\mathbf{u}} \cdot (\delta p\, \mathbf{n} - \mu\, \mathbf{n} \cdot \nabla \delta \mathbf{u}) \,ds \;dt
= \int_0^T \delta D \,dt
\end{equation}
where $S$ denotes the walls over which the drag is calculated.

\end{document}